\documentclass[a4paper,twocolumn,%tightenlines,
english,aps,pre,floatfix,superscriptaddress,showpacs,nofootinbib]{revtex4-1}
\usepackage[T1]{fontenc}
\usepackage[latin1]{inputenc}
\usepackage{amsmath}
\usepackage{babel}
\usepackage{graphics}
\usepackage{amssymb}
\usepackage{dcolumn}


\begin{document}
\title
{Statistics of current activity fluctuations in asymmetric flow with exclusion}
\author {R. B. \surname{Stinchcombe}}
\email{r.stinchcombe1@physics.ox.ac.uk}
\affiliation{Rudolf Peierls Centre for Theoretical Physics, University of
Oxford, 1 Keble Road, Oxford OX1 3NP, United Kingdom}
\author {S. L. A. \surname{de Queiroz}}
\email{sldq@if.ufrj.br}
\affiliation{Rudolf Peierls Centre for Theoretical Physics, University of
Oxford, 1 Keble Road, Oxford OX1 3NP, United Kingdom}
\affiliation{Instituto de F\'\i sica, Universidade Federal do
Rio de Janeiro, Caixa Postal 68528, 21941-972
Rio de Janeiro RJ, Brazil}

\date{\today}

\begin{abstract} 
We consider steady-state current activity statistics for the 
one-dimensional totally asymmetric simple exclusion process (TASEP). With the help of 
the known operator algebra (for general open boundary conditions), as well as general 
probabilistic concepts (for the periodic case), we derive and evaluate closed-form 
expressions for the lowest three moments of the probability distribution function. 
These are confirmed, to excellent degree of accuracy, by numerical simulations.
Further exact expressions and asymptotic approximations are provided for probability
distributions and generating functions.
 
\end{abstract}
%05.40.-a Fluctuation phenomena, random processes, noise, and Brownian motion
%02.50.-r Probability theory, stochastic processes, and statistics
%05.70.Fh Phase transitions: general studies
\maketitle
%\tightenlines
 
\section{Introduction} 
\label{intro} 
The one-dimensional totally asymmetric simple 
exclusion process (TASEP) is a biased diffusion process for particles with 
hard-core repulsion~\cite{derr98,sch00,derr93,rbs01,be07}. 
This model exhibits many non-trivial properties, and is considered a paradigm in the 
field of non-equilibrium phenomena. While a large number of exact results are known 
for the steady-state, ensemble-averaged, values of relevant quantities, such as the
systemwide current and global density, the study of the corresponding fluctuations
has proved to be rather more intricate. To quote only a few relevant developments,
exact expressions for the diffusion constant were found 
for systems with periodic (PBC)~\cite{dem93} and open~\cite{dem95} 
boundary conditions (BC);
the full probability distribution function (PDF) of current fluctuations was
similarly considered for both PBC~\cite{dl98} and open~\cite{bd06} BC. 
Very recently, a number
of new results have been found for current fluctuations in systems with open 
BC~\cite{kvo10,gv11,lm11,ess11}.
 
In this paper we investigate steady-state current activity fluctuations in the 
one-dimensional TASEP for both periodic and open boundary conditions. 
It is important to recall at 
the outset that the current {\em activity} is not identical to 
the standard current (although the first moments of the respective distributions
coincide). As explained in detail below, the former quantity is {\em static},
in this sense akin to the instantaneous (local or global) particle density,
while the latter  is a dynamic one.
Our results extend and complement earlier analytic work on the joint
current-density distribution for the TASEP~\cite{ds04,ds05}.

For the problem of flow with exclusion, the time  evolution 
of the $1+1$ dimensional TASEP is the fundamental discrete model.
The particle number $n_\ell$ at lattice site $\ell$ can be $0$ or $1$, 
and the forward hopping of particles is only to an empty adjacent site. 
The current across the bond from $\ell$ to $\ell +1$ depends also
on the stochastic attempt rate, $p_\ell$, associated with it and is thus
given by  $J_{\ell,\ell+1}= p_\ell\,n_\ell (1-n_{\ell+1})$~. 
For the homogeneous case of $p_\ell =p$ which is considered here, 
one can effectively make $p=1$ in numerical simulations, provided 
that the inherent stochasticity of  the process is kept, via, for example, 
random selection of site occupation update~\cite{dqrbs08}. 
This amounts to a trivial renormalization of the time scale, and is the
procedure followed in all numerical work reported in this paper. 

For PBC the main parameter is the fixed density, which is uniform in the
steady state.
With open boundary conditions, one has the following externally-imposed
parameters: the injection (attempt) rate $\alpha$ at the left end,
and the ejection rate $\beta$ at the right one.
The phase diagram in $\alpha$--$\beta$ parameter space is known exactly, as
well as many other steady state 
properties~\cite{derr98,sch00,derr93,rbs01,be07,ds04,ds05,nas02,ess05}.

In the open-boundary case both current and global (position-averaged) particle density
are fluctuating quantities (as opposed to the case with PBC for which the density is fixed).
On the other hand, while continuity
dictates that the average stationary current on every bond must be the same, steady-state
density profiles may be non-uniform~\cite{derr93}. 

For PBC the total (instantaneous) activity $A$ within the system is defined as the number 
of bonds that can facilitate a transition of a particle in the immediate
future. Thus it equals the number of pairs of neighboring sites that 
have a particle to the left and a hole to the right~\cite{ds04,ds05}.
For systems with open BC, an alternative definition includes 
also the injection  and ejection bonds at the system's ends, though these have to be 
weighted by the respective injection and ejection rates, $\alpha$ and $\beta$.
In the following, for open BC we always include the contributions given by the latter bonds, 
unless explicitly stated otherwise. 

Thus for an $L$--site system with PBC, one has:
\begin{equation}
A =\sum_{\ell=1}^{L} n_\ell\,(1-n_{\ell+1})\qquad {\rm (PBC)}\ ,
\label{eq:adefp}
\end{equation}
(with $n_{L+1} \equiv n_1$), while for the open BC case
($L$ sites and $L+1$ bonds, including the injection and ejection ones), one has:
\begin{equation}
A =\alpha\,(1-n_1)+ \sum_{\ell=1}^{L-1} n_\ell\,(1-n_{\ell+1})+\beta\,n_L\quad
{\rm (Open\ BC)}\ .
\label{eq:adef}
\end{equation}

The activity is, therefore, a snapshot of the system
at a given moment in its evolution; in this sense, it is as much of a static quantity 
as, for example, the instantaneous global density. By contrast, the current is a
dynamic object, as it reflects the stochastically-determined  particle 
displacements which actually take place during a unit time interval.  

The detailed balance measure used here for the PBC case, and 
the operator algebra used for open BC, apply to the steady
state~\cite{be07,dem93,ds04,ds05},
thus they naturally yield the activity; although they can be extended to deal
with static measures of dynamic quantities such as the diffusion 
constant~\cite{dem93,dem95} and more general aspects~\cite{rbs01,ss95}, we
do not attempt similar developments here.

The qualitative distinction between activity and current has the consequence, upon the
(properly normalized) moments of the PDFs
of the two quantities, that while their first moments (averages) coincide, higher 
cumulants differ. For $A$ as defined in Eqs.~(\ref{eq:adefp}),~(\ref{eq:adef}) 
one has, respectively:
\begin{equation}
J=\frac{1}{L}\,\langle A\rangle\qquad\qquad {\rm (PBC)}\  ;
\label{eq:a-jp}
\end{equation}
\begin{equation}
J=\frac{1}{L+1}\,\langle A\rangle\qquad{\rm (Open\ BC)}\ ,
\label{eq:a-j}
\end{equation}
where $J$ is the average steady-state current through any bond in the system. 
Equality of averages can be understood by recalling that
successive steady-state snapshots (activity configurations) are generated via 
the intervening particle hoppings, which constitute realizations of the 
system's current. In our simulations we have verified that this property holds, 
to within numerical accuracy, in all cases investigated here.
However, the connection at this level 
is not sufficient to warrant equality of higher moments of the PDFs. 

Here we consider the fluctuations of the {\it global}, {\it instantaneous} 
activity~\cite{ds04,ds05},  
as opposed to the {\it local}, {\it integrated} current which is the subject of
many extant works~\cite{dem95,dl98,bd06,kvo10,gv11,lm11,ess11}. 

Section~\ref{sec:th} below gives the theoretical approach used in this work.
It first addresses open boundary cases with $\alpha+\beta=1$ and then
general $\alpha$, $\beta$, and then treats PBC. 
In Section~\ref{sec:nr} the results of numerical simulations are given,
first for open BC and then for PBC. 
In Sec.~\ref{sec:conc} we provide a global discussion 
of the second and third cumulants of the activity PDFs,
everywhere on the $\alpha-\beta$ plane (for open BC); 
finally, concluding remarks are made.

\section{Theory}
\label{sec:th}

\subsection{Open boundary conditions with $\alpha+ \beta=1$}
\label{sec:a+b1}

For open boundary conditions, steady state configurations are 
typically highly correlated. So when (as in the pioneering
work of Ref.~\onlinecite{derr93}) such configurations are written as strings of variables 
$D$ and $E$ ($D$, $E$ representing particle, vacancy respectively) 
these variables have to be operators, with an algebra ($D+E=DE$) consistent with the 
steady-state properties. In the special case
$\alpha+\beta=1$ the correlations disappear~\cite{derr93}
(configuration probabilities factorize,
corresponding to a product measure) and $D$, $E$ can be taken as $c$-numbers $d$, $e$
respectively equal to the steady state averages $\langle n_\ell \rangle$, 
$\langle 1- n_\ell \rangle$. Consequently, along the $\alpha+\beta=1$ line 
these averages are independent of $\ell$~\cite{derr93}; consistency
with probability normalization ($d+e=1$) and current conservation 
($e\alpha=de=d\beta$) thus implies $e=\beta$, $d=\alpha$. 

The factorization of probabilities for variables on different sites makes this case the
easiest to begin with.  However, different bonds share a site variable if they are adjacent, 
and that makes the calculation of activity statistics beyond the first moment 
non trivial even in this case.

The average activity $\langle A\rangle$ is unaffected by the 
shared variable on adjacent bonds, since it is the average of a sum of terms each of 
which contains independent site variables and so has a factorizing average. So in this case, 
from Eq.~(\ref{eq:adef}) one has:
\begin{equation}
\langle A\rangle =(L+1)\,\alpha \beta\qquad\quad (\alpha+\beta=1)\ .
\label{eq:aava+b1}
\end{equation} 

Already in the second moment, $\langle A^2\rangle$, the effect of correlations are present,
making it different from $\langle A\rangle^2$. 
To obtain any moment, and also the probability distribution,
we consider the generating function
\begin{equation}
\langle e^{\lambda\,A}\rangle = \langle e^{\lambda\alpha\,(1-n_1)}\, 
\left(\prod_{\ell=1}^{L-1} e^{\lambda\,n_\ell\,(1-n_{\ell+1})}\right)\,e^{\lambda\beta\,
n_L}\,\rangle\ .
\label{eq:genfunc}
\end{equation}

Logarithmic derivatives of $\langle e^{\lambda\,A}\rangle$  give the cumulants $C_n$
of $A$. The probabilities  $P(A_i)$ of the discrete possible outcomes $A_i$ for $A$
are the coefficients in the terms
proportional to $e^{\lambda\,A_i}$ in the expansion of the generating function.

The average in Eq.~(\ref{eq:genfunc}) does not factorize because of the correlation 
of adjacent bond variables. The following method, involving a transfer matrix along the 
chain, easily handles that.

Defining $\zeta_\ell \equiv 1-n_\ell$,
since each $\zeta_\ell$  takes values $(0,1)$ with respective probabilities
$(\alpha,\beta)$, the average in  Eq.~(\ref{eq:genfunc}) can be obtained by summing 
over the possible configurations of each variable 
in turn, starting, say, with $\ell=1$ and working along the chain to $\ell=L$. 
Using sums over the $\zeta_\ell$ variables, with their weights $p(\zeta_\ell)$ ($= \alpha$ 
or $\beta$), 
the first step involves $\Sigma_{\zeta_1=0,1} p(\zeta_1)\, e^{\lambda \alpha\,\zeta_1}\,
e^{\lambda\,(1-\zeta_1)\,\zeta_2} =
\alpha\,e^{\lambda\,\zeta_2} +\beta\,e^{\lambda \alpha} $.
This result can be written in the
linear form $(b_2 + c_2\,\zeta_2)$ where $b_2=\alpha +\beta e^{\lambda \alpha}$
and $c_2 =\alpha\,(e^{\lambda}-1)$. 
The sums over the next variables
then produce similar forms, 
where the relationship between successive ones is provided by
\begin{equation}
b_{\ell+1} + c_{\ell+1}\,\zeta_{\ell+1} = \sum_{\zeta_\ell=1,0}p(\zeta_\ell) 
e^{\lambda (1-\zeta_\ell)\zeta_{\ell+1}}(b_\ell + c_\ell \zeta_\ell)\ .
\label{eq:recdef}
\end{equation}

This is the same as the action of the transfer matrix 
\begin{equation}
T= \begin{pmatrix} \alpha+\beta & \beta \\ \alpha\gamma & 0 \end{pmatrix}\ ,
\label{eq:tdef}
\end{equation}
on the vector $\begin{pmatrix} b_\ell \\ c_\ell \end{pmatrix}$, where $\gamma=
e^\lambda -1$.

The consequence is 
\begin{eqnarray}
\langle e^{\lambda\,A}\rangle =\sum_{\zeta_L} p(\zeta_L)\,\left(b_L+c_L\,\zeta_L
\right)\,e^{\lambda\beta\,(1-\zeta_L)}=
\nonumber \\
=\begin{pmatrix} \alpha\,e^{\lambda\beta}\!+\!\beta &\ \beta \end{pmatrix}\,
T^{L-1}\,\begin{pmatrix} b_1 \\ c_1 \end{pmatrix}\ ,
\label{eq:genfunc2}
\end{eqnarray}
where, from $e^{\alpha\zeta_1}=b_1+c_1\,\zeta_1$, $b_1=1$, $c_1=e^{\lambda\alpha}
-1$.

This is readily evaluated using the representation in which $T$ is diagonal, 
with the result:
\begin{equation}
\langle e^{\lambda\,A}\rangle = a_+\,\mu_+^{L-1}+a_-\,\mu_-^{L-1}\ ,
\label{eq:ev12}
\end{equation}
where
\begin{equation}
\mu_\pm =\frac{1}{2}\left[1 \pm \sqrt{(\alpha-\beta)^2+4\alpha\beta\,e^\lambda}\right] 
\equiv \frac{1}{2}\left[1 \pm \varphi \right]
\label{eq:mupm}
\end{equation}
are the two eigenvalues of $T$, and the coefficients $a_+$, $a_-$ are also known
functions of $\alpha$, $\beta$, $\gamma$ (given in Appendix~\ref{sec:app1}).

This result provides the $n$th cumulant $C_n$ of $A$ via
\begin{equation}
C_n = \left(\frac{\partial}{\partial\lambda}\right)^n\,\ln \langle e^{\lambda\,A}\rangle
\bigg|_{\lambda=0}\ .
\label{eq:cumdef}
\end{equation}
From Eqs.~(\ref{eq:ev12}) and ~(\ref{eq:mupm}),  for large $L$ one has:
\begin{equation}
\ln \langle e^{\lambda\,A}\rangle  \approx L\,\ln \mu_+ + {\cal O}(1)\ .
\label{eq:largeL}
\end{equation}
Consequently the asymptotic large-$L$ values of the first few cumulants are:
\begin{equation}
C_1= \langle  A\rangle \approx L\,\frac{\partial}{\partial \lambda} \ln \mu_+
\bigg|_{\lambda=0}=L\,\alpha\beta + {\cal O}(1)\ ;
\label{eq:c1}
\end{equation}
\begin{equation}
C_2= \langle A^2 \rangle -\langle A\rangle^2 \approx
L\,\alpha\beta\,\left[1-3\alpha\beta\right]+ {\cal O}(1)\ ;
\label{eq:c2}
\end{equation}
\begin{eqnarray}
C_3= \langle A^3 \rangle -3\langle A\rangle\langle A^2 \rangle 
+2 \langle A\rangle^3 \approx\qquad\qquad\qquad \nonumber \\
\approx L\,\alpha\beta\,\left[1-9\alpha\beta+20(\alpha\beta)^2\right]+ {\cal O}(1)
\qquad\ .  
\label{eq:c3}
\end{eqnarray}
Thus the skewness~\cite{numrec} $S \equiv C_3/C_2^{3/2}$ 
is of order $L^{-1/2}$ for large $L$.

The special case $\alpha=\beta=1/2$ is simple and instructive. There the above 
leading large-$L$
results for $C_3$ and skewness reduce to zero, which turns out to be exactly true for any 
$L$: for $\alpha=\beta=1/2$ the coefficient $a_-$ vanishes and the full result from  
Eqs.~(\ref{eq:ev12}) and~(\ref{eq:mupm}) (exact for any L) is:
\begin{equation}
\langle e^{\lambda\,A}\rangle = e^{\lambda/2}\,\left[\frac{1}{2}(1+
e^{\lambda/2})\right]^{L-1}\ .
\label{eq:gfab05}
\end{equation}
From this it is easy to verify that $C_3=S \equiv 0$, and that $C_1=(L+1)/4$,
$C_2=(L-1)/16$.

The corresponding general $L$ exact results obtained from Eqs.~(\ref{eq:ev12}) 
and~(\ref{eq:mupm}) for any $\alpha$, $\beta$ with $\alpha+\beta=1$
are given in Appendix~\ref{sec:app1}, together with validity conditions for 
the large-$L$ results in Eqs.~(\ref{eq:c1})--(\ref{eq:c3}).

For $\alpha=\beta=1/2$ the generating function result, Eq.~(\ref{eq:gfab05}) 
can be expanded in the form
\begin{equation}
\langle e^{\lambda\,A}\rangle = 2^{-(L-1)}\sum_{n=0}^{L-1} {_{L-1}}C_{n}\, 
e^{\frac{\lambda}{2}(n+1)}\ ,
\label{eq:ab05ex}
\end{equation}
corresponding to a binomial distribution of probabilities $P(A_m=m/2)=2^{-(L-1)}\,
{_{L-1}}C_{\,{m-1}}$
for the possible outcomes $m/2$, $1 \leq m \leq L$,  of the activity.

For general $\alpha$, $\beta$ (still with $\alpha+\beta=1$) the possible outcomes for
$A$ will
obviously be linear combinations of $\alpha$, $\beta$, and integers. Their distribution,
resulting from Eqs.~(\ref{eq:ev12}) and~(\ref{eq:mupm}), has 
(for $\alpha \neq \beta$, where $a_-$ is non-vanishing) a "discreteness alternation". 
This is 
a direct consequence of Eqs.~(\ref{eq:ev12}) 
and~(\ref{eq:mupm}), as can be seen from
the following alternative form: 
\begin{equation}
\langle e^{\lambda\,A}\rangle = 2^{-(L-1)}\,\sum_{n=0}^{L-1} {_{L-1}}C_n\,\varphi^n\left[
a_+ +(-1)^n\,a_-\right]\ .
\label{eq:mualt}
\end{equation}   

We conclude this part on activity statistics for $\alpha+\beta=1$ by briefly commenting on
local fluctuations. The product measure makes these trivial for this case. So the injection
current activity (from $\alpha \zeta_1$) is $\alpha$ or $0$, with probabilities $\beta$, 
$\alpha$
respectively. The local activity for the ejection current $[\,\beta\,(1- \zeta_L)\,$] and for 
the
internal current $[\,(1-\zeta_\ell)\,\zeta_{\ell+1}=A_{\ell,\ell+1}\,]$ across any specified 
bond similarly have binary distributions, with (outcomes)(respective probabilities) being
$(\beta,0)$ $(\alpha,\beta)$ and $(1,0)$ $(\alpha \beta,1-\alpha\beta)$, respectively. 
So for example 
$\langle A_{\ell,\ell+1} \rangle =\alpha\beta$, 
$\langle A_{\ell,\ell+1} \rangle^2=\langle A_{\ell,\ell+1} \rangle^3 =\alpha\beta^3$, 
independent of $\ell$, and the
consequent skewness is $(\beta-\alpha)(\alpha\beta)^{-1/2}$.

\subsection{Open boundary conditions with general $\alpha$, $\beta$}
\label{sec:gen_ab}

\subsubsection{Moments of the current activity} 
\label{sec:gen_mom}

For general $\alpha$, $\beta$ one must make full use of
the steady state operator algebra~\cite{derr93} to take care of the absence 
of product measure. In this subsection we derive expressions for low moments  
of the activity. 

We exploit the representation of configurations, and their
probabilities, using strings of operators $D$ and $E$, representing respectively
particle or vacancy at a site. Configurations are specified using the  variables $\zeta_\ell$ 
introduced in Section~\ref{sec:a+b1}.
The probability of the configuration $\{ \zeta_\ell \}$ for $L$ sites is given, through
the associated operator string $S(\{ \zeta_\ell \})=\prod_{j=1}^L 
\left((1-\zeta_j)\,D+\zeta_j\,E\right)$,
as
\begin{equation}
P(\{ \zeta_\ell \})=\frac{\langle W| S(\{ \zeta_\ell \})\,|V\rangle}{Z_L}\ ,\quad
{\rm with}\ Z_L \equiv \langle W| C^L\,| V \rangle\ .
\label{eq:prob_gen}
\end{equation}
Here, $C \equiv D+E$ represents both possibilities (occupied or vacant) at a given site;
and the vectors $\langle W|$ and $| V \rangle$ are such that
\begin{equation}
\langle W|\,\alpha E =\langle W|\ ,\quad \beta D\,|V\rangle=|V\rangle\ .
\label{eq:wv}
\end{equation}
Eqs.~(\ref{eq:wv}) follow from the balance between 
the current $\Lambda \equiv DE$ across any internal bond
and the injection or ejection current (at the left or right boundary), 
using the operator algebra steady-state relation $\Lambda = C$ .
Using Eq.~(\ref{eq:adef}), the $n$th moment of the total activity is given by
\begin{eqnarray}
\langle A^n \rangle=\sum_{\{\zeta_j\}}(\alpha\zeta_1 + \sum_{\ell=1}^{L-1}
\left(1-\zeta_\ell) \zeta_{\ell+1}+\beta (1-\zeta_L)\right)^n\, \times \nonumber \\
\times\,
\frac{\langle W| \prod_{j=1}^L ((1-\zeta_j)\,D+\zeta_j\,E)\,|V\rangle}{Z_L}\ .\hskip1truecm
\label{eq:a^n}
\end{eqnarray}
Consider, for example, the simple case $n=1$. There the contribution from 
$(1-\zeta_\ell)\zeta_{\ell+1}$ is non-zero only for $1-\zeta_\ell=1$, $\zeta_{\ell+1}=1$
whose associated probability is (using $\Lambda \equiv DE=C$)
\begin{equation}
\frac{\langle W| C^{\ell-1}\,DE\,\,C^{L-(\ell+1)}|V\rangle}{Z_L} =
\frac{Z_{L-1}}{Z_L}\ .
\label{eq:exn1}
\end{equation}
Correspondingly, using Eq.~(\ref{eq:wv}) the non-zero contributions
$\alpha$ from $\alpha\zeta_1$ and $\beta$ from $\beta(1-\zeta_L)$ occur
with respective probabilities $\alpha^{-1}\,Z_{L-1}/Z_L$ and
$\beta^{-1}\,Z_{L-1}/Z_L$, making
\begin{equation}
\langle A \rangle = (L+1)\,\frac{Z_{L-1}}{Z_L}\ .
\label{eq:ava_gen}
\end{equation}
This is the standard result generalizing Eq.~(\ref{eq:c1}) for $\alpha+\beta=1$,
where $Z_L=(\alpha\beta)^{-L}$.

As in Section~\ref{sec:a+b1}, complications arise in higher moments from
the shared variable in adjacent bonds. For $\alpha+\beta=1$ the transfer-matrix
approach overcame these. That method is ruled out by the non-commuting variables
in the present general case. Instead the following direct method can be used 
for low $n$.

For $n=2$, Eq.~(\ref{eq:a^n}) includes the product of bond terms 
$\left(\sum_{\ell=1}^{L-1}(1-\zeta_\ell) \zeta_{\ell+1}\right)\,\left(\sum_{m=1}^{L-1}
(1-\zeta_m)\zeta_{m+1}\right)$. For $m-\ell \neq 0,1$ in the double sum the sites involved
are distinct, which simplifies the averages occurring. For example,
for $m>\ell+1$ the average is
\begin{equation}
\frac{\langle W| C^{\ell-1}\,DE\,\,C^{\,m-\ell-2}\,DE\,C^{L-(m+1)}|V\rangle}{Z_L} =
\frac{Z_{L-2}}{Z_L}\ .
\label{eq:exn2}
\end{equation}
This result applies for any $\ell$, $m$ satisfying $m>\ell+1$, or $\ell>m+1$
(altogether $(L-2)(L-3)$ terms of the double sum). For each of the $L-1$
terms with $\ell=m$ the product of bond variables is 
$\left((1-\zeta_\ell)\zeta_{\ell+1}\right)^2=(1-\zeta_\ell)\zeta_{\ell+1}$, having average  
$\langle W| C^{\ell-1}\,\Lambda\,\,C^{L-(\ell+1)}|V\rangle /Z_L=Z_{L-1}/Z_L$.

For the $2(L-2)$ terms with $\ell=m \pm 1$ the product of variables includes both a 
$1-\zeta$ and a $\zeta$ for the shared site, whose product is zero.
Hence, for the part $A^\prime=\sum_{\ell=1}^{L-1}(1-\zeta_\ell)\zeta_{\ell+1}$
of the activity coming from just the internal bonds,
\begin{equation}
\langle {A^\prime}^2\rangle =(L^2-5L+6)\,\frac{Z_{L-2}}{Z_L}+(L-1)\,\frac{Z_{L-1}}{Z_L}\ .
\label{eq:ap2av}
\end{equation}  
Including also the injection and ejection contributions gives
\begin{eqnarray}
\langle A^2\rangle=\langle \left(\alpha\zeta_1 +A^\prime + 
\beta(1-\zeta_L)\right)\rangle^2=\hskip1.5truecm\nonumber\\
\langle {A^\prime}^2\rangle +
\langle \alpha^2\zeta_1+2\alpha\beta\zeta_1(1-\zeta_L)
+\beta^2(1-\zeta_L)\rangle +\hskip1.5truecm\nonumber\\
+\langle2\left(\alpha\zeta_1+\beta(1-\zeta_L)\right)A^\prime\rangle = 
\hskip1.5truecm\nonumber\\
=\langle {A^\prime}^2\rangle+(\alpha+\beta)\,\frac{Z_{L-1}}{Z_L} + 
2\,\frac{Z_{L-2}}{Z_L} +2\Gamma\ ,\hskip1.2truecm
\label{eq:a2av}
\end{eqnarray}
where
\begin{eqnarray}
\Gamma=\langle\big\{\alpha\zeta_1(1-\zeta_1)\zeta_2+\alpha\zeta_1\,\sum_{\ell=2}^{L-1}
(1-\zeta_\ell)\zeta_{\ell+1}+\nonumber\\
+(1-\zeta_{L-1})\zeta_L\,\beta(1-\zeta_L) +\beta(1-\zeta_L)\,
\sum_{\ell=1}^{L-2}(1-\zeta_\ell)\zeta_{\ell+1}\big\}\rangle= \nonumber \\
=2(L-2)\,\frac{Z_{L-2}}{Z_L}\ .\hskip1.truecm
\label{eq:a2av2}
\end{eqnarray}

The case $n=3$ is in principle a straightforward generalization of the above. However,
the $\langle {A^\prime}^3\rangle$ part of it is considerably more complicated. Details 
of the evaluation are given in Appendix~\ref{sec:app2}.

The collected results for the first three moments of both $A$ and $A^\prime$ are: 
\begin{eqnarray}
\mathit{n=1}:\hskip1.5truecm
\langle A \rangle =(L+1)\,\frac{Z_{L-1}}{Z_L}\hskip2truecm\nonumber\\ 
\langle A^\prime \rangle =(L-1)\,\frac{Z_{L-1}}{Z_L}\hskip2truecm
\label{eq:momn1}
\end{eqnarray} 
\begin{eqnarray}
\mathit{n=2}:\hskip6.5truecm\nonumber\\
\langle A^2 \rangle =L(L-1)\,\frac{Z_{L-2}}{Z_L}+(L-1+\alpha+\beta)\,\frac{Z_{L-1}}{Z_L}
\quad\nonumber\\ 
\langle {A^\prime}^2 \rangle =(L-2)(L-3)\,\frac{Z_{L-2}}{Z_L}+(L-1)\,\frac{Z_{L-1}}{Z_L}
\qquad
\label{eq:momn2}
\end{eqnarray} 
\begin{eqnarray}
\mathit{n=3}:\hskip7.25truecm\nonumber\\
\langle A^3 \rangle =(L-1)(L-2)(L-3)\,\frac{Z_{L-3}}{Z_L}+\hskip3truecm
\nonumber\\ +3(L-1)(L-2+\alpha+\beta)
\,\frac{Z_{L-2}}{Z_L}+\hskip2.5truecm \nonumber\\ 
+(L-1+\alpha^2+\beta^2)\,\frac{Z_{L-1}}{Z_L}\hskip3.78truecm
\nonumber\\ 
\langle {A^\prime}^3 \rangle =(L-3)(L-4)(L-5)\,\frac{Z_{L-3}}{Z_L}+\hskip3truecm
\nonumber\\ +3(L-2)(L-3)\,\frac{Z_{L-2}}{Z_L}+(L-1)\,\frac{Z_{L-1}}{Z_L}\ .
\hskip1.5truecm
\label{eq:momn3}
\end{eqnarray} 
The functions $Z_L$ are available from Ref.~\onlinecite{derr93}, or by using the
generating function method of Refs.~\onlinecite{ds04,ds05}. 
The general expression is:
\begin{equation}
Z_L=\sum_{\ell=1}^L \frac{\ell\,(2L-\ell-1)!}{L!\,(L-\ell)!}\,
\sum_{k=0}^\ell \alpha^{-k}\,\beta^{\,k-\ell}\ .
\label{eq:za-bgen}
\end{equation}

Eq.~(\ref{eq:za-bgen}) reduces to simple closed form expressions in the cases:
\begin{eqnarray}
{\rm (i)}\ \alpha+ \beta=1:\qquad Z_L=(\alpha\beta)^{-L}\hskip2.8truecm
\label{eq:znsimpa}\\
{\rm (ii)}\ \alpha=\beta=1:\qquad Z_L=\frac{(2L+2)!}{(L+2)!\,(L+1)!}\ .\hskip1truecm
\label{eq:znsimpb}
\end{eqnarray}
From Eqs.~(\ref{eq:momn1})--(\ref{eq:momn3}) and~(\ref{eq:znsimpb}), 
one can show that, for $\alpha=\beta=1$ and large $L$,  the skewness $S$ 
(of the distribution for $A$) varies as 
\begin{equation}
S=9 L^{-5/2}\,(1+{\cal O}(L^{-1}))\qquad (\alpha=\beta=1)\ .
\label{eq:skab1}
\end{equation}

In Section~\ref{sec:nr-obc} below we describe numerical simulations performed 
for the cases covered by Eqs.~(\ref{eq:znsimpa}) and~(\ref{eq:znsimpb}).
We return to the general $\alpha$, $\beta$ case, using Eq.~(\ref{eq:za-bgen}), in 
Section~\ref{sec:conc}.

\subsubsection{The activity distribution - I}
\label{sec:actdist1}

This section is concerned with the internal bond activity distribution, 
for general $\alpha$, $\beta$. We again use the internal bond current 
activity operator
$\Lambda=DE$, and $C=D+E$, which represents both possible single-site
configurations.

For $n$ successive sites, $C^n$ represents all possible configurations;
we generalize this by representing by $(C^n)$ the subset which contains
no $\Lambda$'s. 
Hereafter in this subsection the round bracket pair $(\quad)$, when enclosing a string 
of $C$'s, or a linear combination of such strings, will represent the formal 
property of picking out the subset containing no $\Lambda$'s.
Then, by definition, $(C^{r_0})\,\Lambda\,(C^{r_1})\,\Lambda\,
(C^{r_2})\,\Lambda\,\cdots\,(C^{r_{n-1}})\,\Lambda\,(C^{r_n})$ contains
many configurations, all having $n$ active internal bonds (total internal activity
$A^\prime=n$), from the $n$ $\Lambda$'s in a run of $L=2n+\sum_{m=0}^n r_m$ sites.
Then,
\begin{equation}
{\cal S}_{L,A^\prime}\{(C^r)\}=\sum_{r_o \geq 0} \cdots \sum_{r_{A^\prime} \geq 0}
(C^{r_0})\,\Lambda\,\cdots\,(C^{r_{A^\prime-1}})\,\Lambda\,
(C^{r_{A^\prime}})\ ,
\label{eq:sdef}
\end{equation}
with $r_0+r_1+\cdots+r_{A^\prime}=L-2A^\prime$, 
contains all configurations with internal activity
$A^\prime$ in an $L$--site system. So, in such a system the probability of
internal current activity $A^\prime$ is
\begin{equation}
P_L(A^\prime) = Z_L^{-1}\,\langle W |\,{\cal S}_{L,A^\prime}\{(C^r)\}\,|V\rangle\ .
\label{eq:pla}
\end{equation}
From their definitions it follows that $(C^n)$ and ${\cal S}_{n,A^\prime}$
are related by
\begin{equation}
(C^n)=C^n-\sum_{A^\prime=1}^{[\,\frac{n}{2}\,]}{\cal S}_{n,A^\prime}\ ,
\label{eq:cvss}
\end{equation}
where $[\,X\,]$ denotes the integer part of $X$~. The steady state
algebra relation $\Lambda=C$ then makes $(C^n)$ a function of $C$ only, so no
lack-of-commutation difficulties arise and one finds that ${\cal S}_{L,A^\prime}$
is the coefficient of $\gamma^{L-2A^\prime}$ in the expansion of
$C^{A^\prime}\,\left(\frac{1}{1-\gamma C}\right)^{A^\prime+1}$. 
That makes ${\cal S}_L \equiv \sum_{A^\prime=1} {\cal S}_{L,A^\prime}$
the coefficient of $\gamma^L$ in $S(\gamma)$, where
\begin{equation}
S(\gamma)= \left(\frac{1}{1-\gamma C}\right)^2\,\gamma^2 C\,\left[1 -
\gamma^2 C\,\left(\frac{1}{1-\gamma C}\right)\right]^{-1}\ .
\label{eq:s_gamma}
\end{equation}
Then, consideration of $\sum_{n=0}^\infty \gamma^n\,{\cal S}_n$ 
eventually gives the following reduction of the formal operation
defined by $(\quad)$:
\begin{equation}
\left(\frac{1}{1-\gamma C}\right)= \frac{1}{1-C(\gamma-\gamma^2)}\ .
\label{eq:fdef}
\end{equation}
Hence, ${\cal S}_{L,A^\prime}$ is the coefficient of $\gamma^{L-2A^\prime}$ in
\begin{equation}
S^\prime(\gamma,A^\prime)=C^{A^\prime}\,\left[1-C(\gamma-\gamma^2)\right]^{-(A^\prime+1)}\ .
\label{eq:coeff}
\end{equation}
The required coefficient can be found by the expansion of the function of $C$
in powers of $(\gamma-\gamma^2)$, and then using the binomial expansion of each
power of $(1-\gamma)$ occurring. Taking the matrix element $\langle W|\,\cdots\,
|V\rangle$ of the resulting $C$--dependent coefficient and using the
definition of $Z_N$ then provides the following exact result for the
internal activity distribution:
\begin{equation}
P_L(A^\prime)=\frac{1}{Z_L}\!\sum_{s=0}^{\left[\frac{L-2A^\prime}{2}\right]}\!Z_{L-A^\prime-s}
(-1)^s\frac{(L-A^\prime-s)!}{A^\prime!s!(L-2A^\prime-2s)!}\ .
\label{eq:papfinal}
\end{equation}
An alternative expansion strategy from Eq.~(\ref{eq:coeff}) onward is available
for the special case $\alpha+\beta=1$. Because the operators reduce to $c$-numbers
in this case, $C$ in that Equation becomes $(\alpha\beta)^{-1}$, so 
$[1-C(\gamma-\gamma^2)]$ becomes $(\alpha\beta)^{-1}(\gamma-\gamma_+)(\gamma-\gamma_-)$
where $\gamma_\pm=\frac{1}{2}(1 \pm \sqrt{1-4\alpha\beta})=\alpha$, $\beta$
(the same as the eigenvalues $\mu_\pm(\lambda)$ of the transfer matrix of 
Section~\ref{sec:a+b1}, at $\lambda=0$). 

In particular, $\gamma_+=\gamma_-=\frac{1}{2}$ for $\alpha=\beta=\frac{1}{2}$, so it
follows that
\begin{equation}
P_L(A^\prime)=2^{-L}\,_{L+1}C_{2A^\prime+1}\ .
\label{eq:papab05}
\end{equation}
For large $L$,  where boundary current contributions to  $A$ are less by a 
factor ${\cal O}(1/L)$ from the total internal bond current contributions
which comprise $A^\prime$, Eq.~(\ref{eq:papab05}) becomes the same as
the result from Eq.~(\ref{eq:ab05ex}). 

Though the method just presented applies for any $\alpha$, $\beta$, it is
not suitable for inclusion of boundary current contributions. We briefly
proceed next to a more sophisticated approach to the current activity
distribution which is in no such way limited, and is related to previous 
work~\cite{ds04,ds05} on the joint distribution for number and current
activity.   

\subsubsection{The activity distribution - II; Generating functions}
\label{sec:actdist2}

The method of Refs.~\onlinecite{ds04,ds05} represented any microscopic configuration
of $N$ particles with internal activity $A^\prime$ in the $L$--site open
boundary TASEP as a sequence of $A^\prime$ objects of form
$D^{p_j}\,E^{h_j}$, $p_j,\,h_j \geq 1$, with $h_0 \geq 0$ $E$'s  to the
left and $p_0 \geq 0$ $D$'s to the right because of the possible injection and 
ejection. The operator string representation for the set of all such 
configurations is obtained by summing over all $\{h_j,\,p_j\}$ as specified
above, subject to the additional constraints coming from specified $L$, $N$,
and $A^\prime$:
\begin{equation}
\sum_{j=0}^{A^\prime}(h_j+p_j)=L\ ,\qquad\sum_{j=0}^{A^\prime}p_j=N\ .
\label{eq:constr}
\end{equation} 
$h_0>0$ (or $p_0>0$) corresponds to unit injection (or ejection) current activity,
accounting for any difference between $A^\prime$ and $A$. The steady state
measure is then obtained from the $\langle W|\,\cdots\,|V\rangle$ matrix element
of the operator string. This was reduced in Refs.~\onlinecite{ds04,ds05} using a
generalized version of the original operator algebra of Ref.~\onlinecite{derr93}.

We are now interested in just the activity distribution, for specific $L$,
so the constraint on $N$ drops out and the remaining constraints can be
enforced using an integral representation of a $\delta$--function.
The resulting expression for the total activity probability function $P_L(A)$
can then be reduced using the modified operator algebra of Refs.~\onlinecite{ds04,ds05}.
The result is that $P_L(A)$ is the coefficient of $z^{L-2A}$ in
\begin{equation}
G(\alpha,\beta,A,z)=\frac{\alpha\beta}{Z_L^{\alpha\beta}}\frac{Z_A
^{\alpha^\prime\beta^\prime}}{(z-\alpha)(z-\beta)(1-2z)^{2A}}\ ,
\label{eq:fgen}
\end{equation}
where, in the numerator, $\alpha^\prime=(\alpha-z)/(1-2z)$,
$\beta^\prime=(\beta-z)/(1-2z)$ replace $\alpha$, $\beta$ to obtain
the "renormalized" version from the usual $Z_L \equiv Z_L^{\alpha\beta}$.
Equivalently, the double-integral representation of Refs.~\onlinecite{ds04,ds05}
for the joint distribution $P_L(A,N)$ can be summed over $N$ to provide the
above form.

As well as giving rise to the "renormalized" $Z_A^{\alpha^\prime\beta^\prime}$, the
modified algebra also gives a generating functional for $Z_L^{\alpha\beta}$; that can be
exploited following Ref.~\onlinecite{ds05} to give a direct form for the
generating function $F_L^{\alpha\beta}(\lambda)\equiv\langle e^{\lambda A}\rangle$
as the coefficient of $z^{L+1}$ in
\begin{eqnarray}
U_L^{\alpha\beta}(z,\lambda)= \frac{4\alpha\beta}{Z_L^{\alpha\beta}}\,[\{2\alpha-1+
\sqrt{(1-2z)^2-4z^2\,e^\lambda}\} \times  \nonumber \\
\times \{2\beta-1+\sqrt{(1-2z)^2-4z^2\,e^\lambda}\}]^{-1}\ .\hskip1truecm 
\label{eq:gencoeff}
\end{eqnarray} 
As expected, for $\alpha+\beta=1$ the square bracket simplifies, and in the special 
case $\alpha+\beta=\frac{1}{2}$ it becomes a product of two factors linear in $z$.
From this it is easy to recover Eq.~(\ref{eq:gfab05}) in Section~\ref{sec:a+b1}.

The form obtained from Eq.~(\ref{eq:fgen}) for the distribution function is not
easy to simplify, except in the asymptotic large $L$ limit. There, saddle point 
integrals analogous to those in Refs.~\onlinecite{ds04,ds05} give the activity
distribution around its peak, apart from $A$-independent factors or others
subdominant at large $L$. 

The asymptotic results for all phases can alternatively be obtained by
integrating out the density from the asymptotic joint distribution results
obtained in Refs.~\onlinecite{ds04,ds05}. So, the activity distribution
can be written for large $L$ in the form
\begin{equation}
P_L(A)=\int_0^1d\rho\,\exp [-L\,g( \rho,j)\,]\ ,
\label{eq:plag}
\end{equation} 
where for the low current phase with $\alpha < \beta$
\begin{eqnarray}
g(\rho,j)=2j \ln j+ (\rho-j) \ln (\rho-j)+ \hskip2.0truecm \nonumber \\
+(1-\rho-j) \ln (1-\rho-j)+\hskip2.truecm \nonumber \\
+\rho \ln \left((1-\alpha)/\rho\right)+(1-\rho)\ln \left(\alpha/(1-\rho)\right)
\ ,\hskip1truecm
\label{eq:grhoj}
\end{eqnarray}
and for the maximal current phase $g(\rho,j)$ is of similar form but
without the last two terms in Eq.~(\ref{eq:grhoj}).

The leading large--$L$ approximation to Eq.~(\ref{eq:plag}), provided by the
Laplace method (i.~e., $\exp [-L\,g({\overline \rho},j)\,]$ where
${\overline \rho}(j)$ in $[0,1]$ minimizes $g$) gives, for example, a maximal current 
phase PDF independent of $\alpha$ and $\beta$, and the same as the Laplace result
for the low current phase PDF at $\alpha=\beta=\frac{1}{2}$, which is consistent
there with the general--$L$ result under Eq.~(\ref{eq:ab05ex}). 

Though of theoretical interest, the results for probability distributions
obtained by the methods of the last two sections are of less value for 
detailed comparisons with numerical simulation results than the exact tractable 
expressions provided, for any $L$, in Section~\ref{sec:gen_mom} for low
moments of the activity. See also Section~\ref{sec:conc}.

\subsection{Periodic boundary conditions}
\label{sec:pbct}

For PBC, the general ideas concerning operator strings,
recalled in section~\ref{sec:gen_mom}, are again applicable. 
A number of simplifying features occur, besides having a fixed number
of particles in the system. From the very simple form
of detailed balance equations applying in this case~\cite{dem93}, 
for a given number $M$ of particles all 
steady-state configurations $\cal C$ are equiprobable:
\begin{equation}
P({\cal C})=\frac{1}{_LC_M}\qquad {\rm (PBC)}\ .
\label{eq:equip}
\end{equation}
We introduce the quantity
\begin{equation}
a_n \equiv \frac {(M-n)(L-M-n)}{L-n-1}\ ,
\label{eq:andef}
\end{equation}
for $n=0$, $1$, $2$, which will prove convenient in what follows.

With the notation established in Section~\ref{sec:a+b1}, the average of the
two-site function $(1-\zeta_\ell)\,\zeta_{\ell+1}$ is independent of $\ell$,
and proportional to the number of
combinations in which site $\ell$ is occupied, site $\ell+1$ is vacant,
and the remaining $M-1$ particles are distributed among the remaining $L-2$
sites. So,
\begin{equation}
\langle A \rangle=L\,\langle (1-\zeta_\ell)\,\zeta_{\ell+1} \rangle=
\frac{_{L-2}C_{M-1}}{_LC_M}=a_0 .
\label{eq:momn1p}
\end{equation}

For $\langle A^2\rangle$, similarly to the discussion in Section~\ref{sec:gen_mom},
one needs averages of four-operator products, namely ${\cal X}_{\ell m} \equiv \langle 
(1-\zeta_\ell)\,\zeta_{\ell+1} \,(1-\zeta_m)\,\zeta_{m+1}\rangle$. Three 
possibilities arise:
\smallskip\par\noindent
(i)\ $m=\ell\,[\,L\ {\rm terms}\,]:\ {\cal X}_{\ell \ell}=\langle 
(1-\zeta_\ell)\,\zeta_{\ell+1} \rangle=a_0/L$ [$\,$using $\zeta_\ell^2=\zeta_\ell$, 
etc, and
Eq.~(\ref{eq:momn1p})$\,$];
\smallskip\par\noindent
(ii)\ $m=\ell\pm 1\ [\,2L\ {\rm terms,\ from\ the\ ring\ geometry}\,]:\ 
{\cal X}_{\ell\,\ell\pm 1} =0$ 
[$\,$since $(1-\zeta_i)\,\zeta_i=0\,$]; and
\smallskip\par\noindent
(iii)\ $m \neq\ell$, $\ell\pm 1\ [\,L^2-3L\ {\rm terms}\,]$:  contributions come 
from having sites $\ell$, $m$ occupied, $\ell+1$, $m+1$ vacant, and $M-2$
particles distributed among the remaining $L-4$ sites, so 
${\cal X}_{\ell\,m}=_{L-4}C_{M-2}/_LC_M$.
\smallskip\par\noindent
Adding up (i)--(iii):
\begin{equation}
\langle A^2 \rangle = a_0\,(1+a_1) \ .
\label{eq:momn2p}
\end{equation} 
Details of the evaluation of $\langle A^3 \rangle$ are given in 
Appendix~\ref{sec:app3}. The result is:
\begin{equation}
\langle A^3 \rangle=a_0\,(1+3 a_1+a_1\,a_2)\ .
\label{eq:momn3p}
\end{equation}
Introducing the variables $\rho \equiv M/L$; $\mu \equiv \rho(1-\rho)$; 
$\Lambda \equiv \mu L^2$; and 
$\widetilde{L} \equiv L-1$, Eqs.~(\ref{eq:momn1p})--(\ref{eq:momn3p}) give the cumulants 
of the PDF in a form which evinces the expected particle-hole duality
(invariance under $\rho \leftrightarrow 1 -\rho$):
\begin{equation}
C_1= \frac{\Lambda}{\widetilde{L}}\ ;
\label{eq:c1p}
\end{equation} 
\begin{equation}
C_2= \frac{\Lambda\,[\Lambda - \widetilde{L}]}{\widetilde{L}^2\,(\widetilde{L}-1)}\ ;
\label{eq:c2p}
\end{equation} 
\begin{equation}
C_3= \frac{\Lambda\,
[4\Lambda^2-\Lambda(\widetilde{L}^2 +6\widetilde{L}) + \widetilde{L}^2\,
(\widetilde{L}+2)]}{\widetilde{L}^3\,(\widetilde{L}-1)(\widetilde{L}-2)}\ .
\label{eq:c3p}
\end{equation} 
For large $L$, the skewness $S$ is given, to leading order in $L^{-1}$, by
\begin{equation}
S=\mu^{-1}\,(4\mu-1)\,L^{-1/2}\ ,
\label{eq:spbc1}
\end{equation}
provided that $\mu \neq 1/4$, i.e. $\rho \neq 1/2$. In the latter case, consideration
of higher-order terms shows that
\begin{equation}
S=4\,L^{-5/2}\,\left[1+{\cal O}(L^{-1})\right]\qquad\qquad (\rho=1/2)\ .
\label{eq:spbc2}
\end{equation}

\section{Numerical results}
\label{sec:nr}

\subsection{Introduction}
\label{sec:nr-intro}

We considered lattices with $L=2^m$ sites. We found that
the size range corresponding to $5 \leq m \leq 8$ is generally 
suitable to highlight $L$--dependent 
effects which vanish for $L \to \infty$, and for which we can provide
accurate numerical checks of the theory
developed in Sections~\ref{sec:a+b1},~\ref{sec:gen_mom}, and~\ref{sec:pbct}. 
On the other hand, such sizes are large enough
to prevent lattice discreteness from playing a significant role [$\,$except for the
"discreteness alternation" effect referred to above, in connection with
Eq.~(\ref{eq:mualt}), which for systems with open BC is a general feature 
unless $\alpha=\beta=\frac{1}{2}$ or~$1\,$]. For very low densities with PBC,
for reasons explained in Section~\ref{sec:nr-pbc} below, we extended the 
size range up to $m=11$.

A time step is defined as a set of $L$ sequential update attempts,
each of these according to the following rules: (1) select a site at random;
(2a) if the chosen site is the rightmost one and is occupied, then
(3a) eject the particle from it with probability $\beta$; alternatively,
(2b) if the site is the leftmost one and is empty, then (3b) inject a
particle onto it with probability $\alpha$; finally, if neither (2a) nor (2b)
is true, (2c) if the site
is occupied and its neighbor to the right is empty, then (3c) move the particle. 

Thus, in the
course of one time step, some sites may be selected more than once for examination,
and some may not be examined at all. 

For the various sets of $\alpha$, $\beta$ (for open BC) or values of $\rho$ (for PBC)
considered here [$\,$because of particle-hole
duality, we take only $\alpha \leq \beta$ for the former case,
and density $\rho \leq 1/2$ for the latter$\,$], we
ascertained that, starting from an initial random configuration of occupied sites
(usually with overall density $\rho=1/2$), $n_{\rm in}=4,000$ time steps was long
enough for steady-state flow to be fully established, 
for all system sizes $L \leq 256$. This also applies to the special case of 
systems with PBC, $L \leq 2048$ and very low densities, examined in 
Section~\ref{sec:nr-pbc}.

We collected steady-state current activity data (typically $N_{\rm sam}=10^5$--$10^7$ 
independent
samples) to produce the respective PDFs. Accurate evaluation of 
their moments of $n$-th order, $1 \leq n \leq 3$, involves running $N_{\rm set}$
independent sets of $N_{\rm sam}$ samples each; from the spread among the averaged moments
for the distinct sets, one then estimates the root-mean-square (RMS) deviation of each
relevant quantity. As is well known~\cite{dqrbs96}, such RMS deviations
are essentially independent of $N_{\rm set}$
as long as $N_{\rm set}$ is not too small, and vary as $N_{\rm sam}^{-1/2}$.  
We generally took $N_{\rm set}=10$. 

Our results for the activity PDF are given in terms of 
the reduced variable $x \equiv (A-\langle A\rangle)/L_0$ [$\,L_0= L+1$
for open BC, or $L$ for PBC$\,$] which, as a
consequence of Eqs.~(\ref{eq:a-j}) and~(\ref{eq:a-jp}), is suitable to highlight 
similarities and differences between the fluctuation statistics of activity and those of
the standard current $J$.

\subsection{Open boundary conditions}
\label{sec:nr-obc}

\begin{figure}
{\centering \resizebox*{3.3in}{!}{\includegraphics*{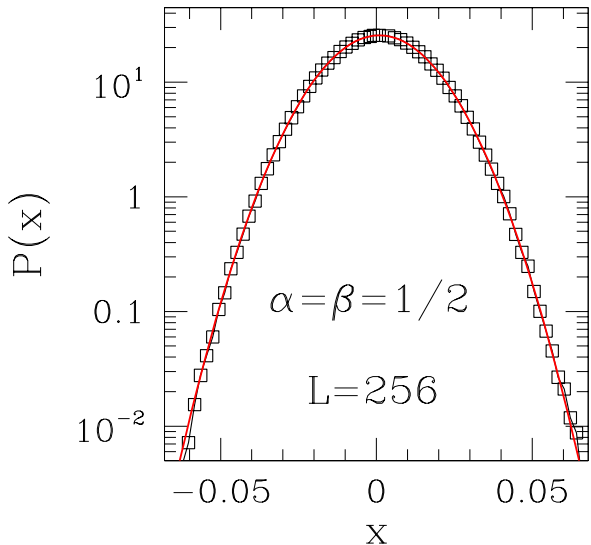}}}
\caption{(Color online) 
PDF for the reduced variable $x \equiv (A-\langle A \rangle)/(L+1)$
for $\alpha=\beta=1/2$, $L=256$ [$\,$see
Eqs.~(\protect{\ref{eq:a-j}}),~(\protect{\ref{eq:aava+b1}})$\,$]. 
The points are from numerical simulations, $N_{\rm sam}=10^6$ samples. 
The full line is the binomial distribution given following
Eq.~(\protect{\ref{eq:ab05ex}}).
} 
\label{fig:ab05}
\end{figure}

Initially we consider the $\alpha+\beta=1$ line.
For the special point $\alpha=\beta=1/2$, the considerations of
Sec.~\ref{sec:a+b1} [$\,$see also Eq.~(\ref{eq:papab05}) in 
Sec.\ref{sec:actdist1}$\,$] indicate that the PDF is a binomial function.
Fig.~\ref{fig:ab05} shows the corresponding numerical simulation data 
for a system with $L=256$ sites, together with the appropriate
binomial distribution (which, for this value of $L$, is indistinguishable
from a Gaussian function of a continuous variable). The fit between the
two sets is indeed excellent.  

\begin{figure}
{\centering \resizebox*{3.3in}{!}{\includegraphics*{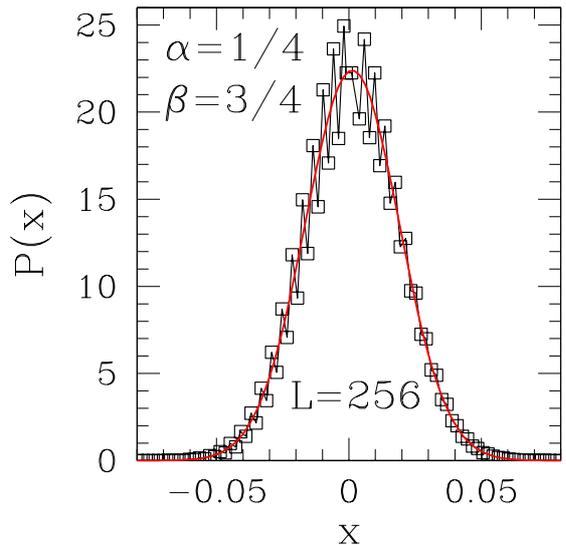}}}
\caption{(Color online) 
PDF for the reduced variable $x \equiv (A-\langle A \rangle)/(L+1)$
for $\alpha=1/4$, $\beta=3/4$, $L=256$ [$\,$see
Eqs.~(\protect{\ref{eq:a-j}}),~(\protect{\ref{eq:aava+b1}})$\,$]. 
The points are from numerical simulations, $N_{\rm sam}=10^6$ samples,
and exhibit the discreteness alternation implied by 
Eq.~(\protect{\ref{eq:mualt}}). 
The full line is a Gaussian fit to the full set of simulation data.
} 
\label{fig:a25b75}
\end{figure}

On the $\alpha+\beta=1$ line, the discreteness alternation effect 
associated with Eq.~(\ref{eq:mualt}) 
is illustrated for $\alpha=1/4$, $\beta=3/4$ in Fig.~\ref{fig:a25b75}. 
To avoid smoothing out of the effect, binning of the variable
 $x=(A-\langle A\rangle)/(L+1)$ must be carefully chosen; for the
simulations reported in Fig.~\ref{fig:a25b75} we used bins of width
$[2(L+1)]^{-1}$.
In order to give higher resolution to the central part of the distribution,
where the sawtooth pattern is quantitatively more significant, a linear 
scale has been used on the vertical axis of the Figure.
The Gaussian fit shown in Fig.~\ref{fig:a25b75} takes into account 
the full set of data, thus averaging over the oscillations.
Remarkably, it produces a rather accurate fitted value, $\Delta=0.0178(3)$
for the corresponding width of the distribution (see Table~\ref{t1} below for
comparison).  

\begin{figure}
{\centering \resizebox*{3.3in}{!}{\includegraphics*{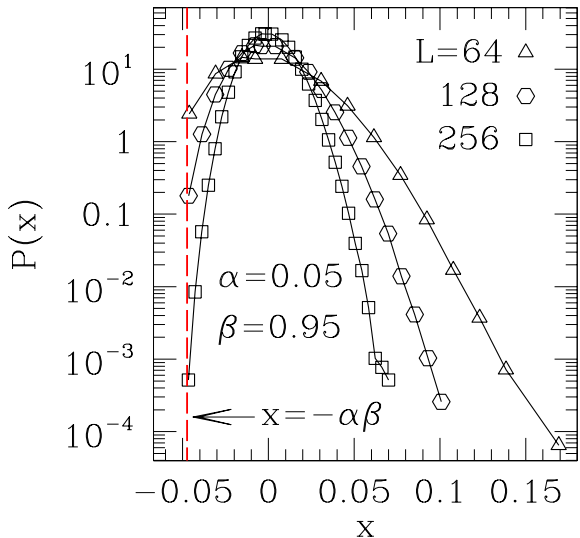}}}
\caption{(Color online) 
PDFs for the reduced variable $x \equiv (A-\langle A \rangle)/(L+1)$
for systems with open boundaries, $\alpha=0.05$, $\beta=1-\alpha$, and sizes $L$ as 
specified [$\,$see Eqs.~(\protect{\ref{eq:a-j}}) and~(\protect{\ref{eq:aava+b1}})$\,$]. 
The points are from numerical simulations, $N_{\rm sam}=10^6$ samples. 
The dashed (red) line corresponds to the minimum allowed activity (see text). 
} 
\label{fig:a05vl}
\end{figure}

Still on the $\alpha+\beta=1$ line, for small $\alpha$ and
large $\beta$ (or, by particle-hole duality, for large $\alpha$ and small 
$\beta$), Eqs.~(\ref{eq:c2}) and~(\ref{eq:c3}) show that for large $L$ the
skewness $S$ is positive and large, proportional to $(\alpha\beta)^{-1/2}$.
The average activity being $\langle A\rangle =(L+1)\alpha\beta$, the minimum allowed 
value of the quantity $x=(A-\langle A\rangle)/(L+1)$ is $x_{\rm min}=-\alpha\beta$, 
i.e., $|x_{\rm min}| \ll 1$ for $\alpha \ll 1$, while positive values
of $x \lesssim 0.5$ are permitted. This is illustrated in Fig.~\ref{fig:a05vl},
where the discreteness alternation has been averaged out by using bins of width 
$(L+1)^{-1}$ for the $x$ variable.
The calculated
values of $S$ for the curves shown are $=0.4471\dots$, $0.3158 \dots$, and
$0.2232\dots$, respectively for $L=64$, $128$, $256$ [$\,$see 
Eqs.~(\ref{eq:momn1})--(\ref{eq:znsimpa})$\,$]. Large, positive skewness
is expected to be a general feature also away from
the $\alpha+\beta=1$ line, anywhere for small $\alpha$, large $\beta$
(and the complementary region at large $\alpha$, small $\beta$). 

\begin{figure}
{\centering \resizebox*{3.3in}{!}{\includegraphics*{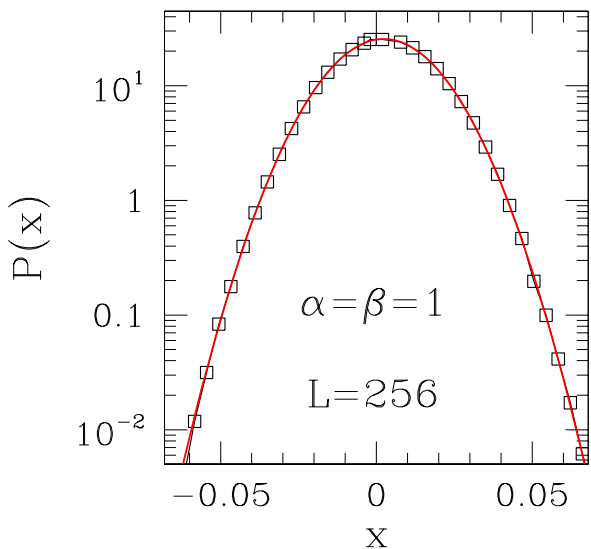}}}
\caption{(Color online) 
PDF for the reduced variable $x \equiv (A-\langle A \rangle)/(L+1)$
for $\alpha=\beta=1$, $L=256$ [$\,$see
Eqs.~(\protect{\ref{eq:a-j}}),~(\protect{\ref{eq:momn1}}), 
and~(\protect{\ref{eq:znsimpb}})$\,$]. 
The points are from numerical simulations, $N_{\rm sam}=10^6$ samples. 
The full line is a Gaussian fit to simulation data.
} 
\label{fig:ab1}
\end{figure}

Away from the $\alpha+\beta=1$ line, we report data for $\alpha=\beta=1$,
where the averaged moments are given by relatively simple expressions
[$\,$see Eq.~(\ref{eq:znsimpb})$\,$]. The activity PDF is shown in
Figure~\ref{fig:ab1}, together with the result of a Gaussian fit to the data.
Visual inspection indicates that the fit is of a similar 
quality to that for the $\alpha=\beta=1/2$ data, displayed in 
Fig.~\ref{fig:ab05}. Indeed, according to 
Eqs.~(\ref{eq:momn1})--(\ref{eq:momn3}), (\ref{eq:znsimpb}), and~(\ref{eq:skab1}), 
although a non-vanishing skewness must
be present in this case, it is quantitatively rather small for large $L$
[$\,$see also Table~\ref{t1}$\,$].

Results from numerical evaluation  of the first three moments of the activity PDF, for the
values of $\alpha$, $\beta$ discussed above, are given 
in Table~\ref{t1}, together with the theoretical predictions summarized 
in Eqs.~(\ref{eq:momn1})--(\ref{eq:znsimpb}). The agreement is very good, except that
accuracy for the skewness is comparatively low. This is especially true for
$\alpha=\beta=1$, in which case 
much narrower error bars would be needed in order to verify the
rather small predicted values.
 
\begin{table}
\caption{\label{t1}
For systems with $L$ sites, and $\alpha$, $\beta$ as specified, $\langle a_L\rangle \equiv 
\langle A \rangle/(L+1)$ is average current activity 
[$\,$normalized as in
 Eq.~(\protect{\ref{eq:a-j}})$\,$]; $\Delta \equiv [\langle a_L^2\rangle -
\langle a_L\rangle^2\,]^{1/2}$; 'Th' stands for results from theory 
[$\,$Eqs.~(\protect{\ref{eq:momn1}})--(\protect{\ref{eq:znsimpb}})$\,$], and 'N' for
results of numerical simulations with $N_{\rm sam}=10^6$, $N_{\rm set}=10$ (see text).
}
\vskip 0.2cm
\begin{ruledtabular}
\begin{tabular}{@{}ccccc}
$L$ & Type  & $\langle a_L \rangle$ & $\Delta$ & Skew  \\
\hline\noalign{\smallskip}
\multicolumn{5}{c}{$\alpha=\beta=1/2$}\\
\hline\noalign{\smallskip}
$64$ &   Th  &   $1/4$      & $\,0.0305279\dots$ & $0$          \\
\ \  &   N   & $0.24993(2)$ & $0.030521(22)$   & $-0.0008(22)$\\
$128$&   Th  &   $1/4$      & $\,0.0218399\dots$ & $0$          \\
\ \  &   N   & $0.24998(2)$ & $0.021850(13)$   & $-0.0005(20)$\\
$256$&   Th  &   $1/4$      & $\,0.0155338\dots$ & $0$          \\
\ \  &   N   & $0.24998(1)$ & $0.015536(14)$   & $\ \ 0.0017(26)$ \\
\hline\noalign{\smallskip}
\multicolumn{5}{c}{$\alpha=1/4$, $\beta=3/4$}\\
\hline\noalign{\smallskip}
$64$ &   Th  &   $3/16$     & $\,0.0351324\dots$ & $\,0.01648\dots$   \\
\ \  &   N   & $0.18744(3)$ & $0.035138(16)$     &  $0.0162(22)$\\
$128$&   Th  &   $3/16$     & $\,0.0250771\dots$ & $\,0.01134\dots$   \\
\ \  &   N   & $0.18747(3)$ & $0.025080(17)$   & $0.0118(22)$\\
$256$&   Th  &   $3/16$     & $\,0.0178161\dots$ & $\,0.00790\dots$   \\
\ \  &   N   & $0.18747(2)$ & $0.017823(13)$   & $0.0071(23)$ \\
\hline\noalign{\smallskip}
\multicolumn{5}{c}{$\alpha=\beta=1$}\\
\hline\noalign{\smallskip}
$64$ &   Th  & $\ \ 0.255814\dots$     & $\,0.0311221\dots$ & $\,0.000091968\dots$   \\
\ \  &   N   & $0.25583(4)$ & $0.031183(20)$     &  $\!\!\!-0.0017(19)$\\
$128$&   Th  & $\ \ 0.252918\dots$     & $\,0.0220529\dots$ & $\,0.000016219\dots$   \\
\ \  &   N   & $0.25293(2)$ & $0.022083(15)$   & $0.0005(27)$\\
$256$&   Th  & $\ \ 0.251462\dots$     & $\,0.0156096\dots$ & $\,0.000002864\dots$   \\
\ \  &   N   & $0.25146(2)$ & $\!\!\!0.015618(8)$   & $0.0008(12)$ \\
\end{tabular}
\end{ruledtabular}
\end{table}   

In order to clarify the latter point, we made long 
runs with $N_{\rm sam}=10^7$ for $L=16$ with $\alpha=\beta=1$,
evaluating both $\langle A^n\rangle$ and $\langle {A^\prime}^n\rangle$  of
Eqs.~(\ref{eq:momn1})--(\ref{eq:momn3}), with the results displayed in 
Table~\ref{t2}. One can see that the numerical results clearly confirm
the theoretical prediction for the skewness, within error bars.

\begin{table}
\caption{\label{t2}
For systems with $L=16$ sites,  $\alpha=\beta=1$, results for total activity: 
(a) including ($\langle A^n \rangle$) and (b) not including  ($\langle 
{A^\prime}^n\rangle$) injection and ejection bond contributions.
 $\Delta \equiv [\langle A^2\rangle -
\langle A\rangle^2\,]^{1/2}$; 'Th' stands for results from theory 
[$\,$Eqs.~(\protect{\ref{eq:momn1}})--(\protect{\ref{eq:momn3}}), 
and~(\protect{\ref{eq:znsimpb}})$\,$], and 'N' for
results of numerical simulations with $N_{\rm sam}=10^7$, $N_{\rm set}=10$ (see text).
}
\vskip 0.2cm
\begin{ruledtabular}
\begin{tabular}{@{}cccc}
 Type  & $\langle A \rangle$ & $\Delta$ & Skew  \\
\hline\noalign{\smallskip}
\multicolumn{4}{c}{(a) $\langle A^n \rangle$}\\
\hline\noalign{\smallskip}
Th  &   $\,4.636363\dots$      & $\,1.042933\dots$ & $\,0.003006\dots$ \\
N   & $4.63655(37)$ & $1.04286(25)$   & $0.00295(69)$\\
\hline\noalign{\smallskip}
\multicolumn{4}{c}{(b) $\langle {A^\prime}^n \rangle$}\\
\hline\noalign{\smallskip}
Th  & $\,4.090909\dots$ & $\,0.982518\dots$ & $\,0.009706\dots$  \\
N   & $4.09113(34)$   & $0.98246(12)$     & $0.00959(34)$\\
\end{tabular}
\end{ruledtabular}
\end{table}   

\subsection{Periodic boundary conditions}
\label{sec:nr-pbc}

For systems with PBC, the $\rho \leftrightarrow 1-\rho$ duality 
exhibited in Eqs.~(\ref{eq:c1p})--(\ref{eq:c3p})
ensures that all relevant aspects can be investigated by considering
only, e. g., $\rho \leq 1/2$.
We evaluated the first three moments of the activity PDF
for densities $\rho=1/2$ and $1/4$, and varying system sizes $L$. Results are given 
in Table~\ref{t3}, together with the theoretical predictions summarized 
in Eqs.~(\ref{eq:momn1p})--(\ref{eq:momn3p}). Similarly to the open boundary case, the 
agreement is very 
good, except that accuracy for the skewness is comparatively low. For $\rho=1/2$,
in particular, comparison of simulational data with predicted values of
skewness [$\,$see Eq.~(\ref{eq:spbc2})$\,$] is difficult because the 
latter are rather small.
\begin{table}
\caption{\label{t3}
For systems with PBC, $L$ sites, and $\rho$ as specified, $\langle a_L\rangle \equiv 
\langle A \rangle/L$ is average current activity 
[$\,$normalized as in
 Eq.~(\protect{\ref{eq:a-jp}})$\,$]; $\Delta \equiv [\langle a_L^2\rangle -
\langle a_L\rangle^2\,]^{1/2}$; 'Th' stands for results from theory 
[$\,$Eqs.~(\protect{\ref{eq:momn1p}})--(\protect{\ref{eq:momn3p}})$\,$], and 'N' for
results of numerical simulations with $N_{\rm sam}=10^6$, $N_{\rm set}=10$ (see text).
}
\vskip 0.2cm
\begin{ruledtabular}
\begin{tabular}{@{}ccccc}
$L$ & Type  & $\langle a_L \rangle$ & $\Delta$ & Skew  \\
\hline\noalign{\smallskip}
\multicolumn{5}{c}{$\rho=1/2$}\\
\hline\noalign{\smallskip}
$64$ &   Th  & $\ \ 0.253968\dots$ & $\,0.0312461\dots$ & $\,0.000130123\dots$   \\
\ \  &   N   & $0.25398(2)$ & $\!0.031261(20)$     &  $-0.0005(19)$\\
$128$&   Th  & $\ \ 0.251969\dots$ & $\, 0.0220964\dots$ & $\,0.000022272\dots$   \\
\ \  &   N   & $0.25197(2)$ & $\!0.022095(11)$   & $0.0004(17)$\\
$256$&   Th  & $\ \ 0.250980\dots$ & $\,0.0156249\dots$ & $\,0.000003875\dots$   \\
\ \  &   N   & $0.25099(1)$ & $\!0.015629(10)$   & $0.0001(26)$ \\
\hline\noalign{\smallskip}
\multicolumn{5}{c}{$\rho=1/4$}\\
\hline\noalign{\smallskip}
$64$ &   Th  & $\ \ 0.190476\dots$ & $\,0.0231771\dots$ & $-0.17946\dots$ \\
\ \  &   N   & $0.19048(2)$ & $\!0.023178(11)$   & $\!-0.1799(13)$\\
$128$&   Th  & $\ \ 0.188976\dots$ & $\,0.0164837\dots$ & $-0.12226\dots$ \\
\ \  &   N   & $ 0.18898(1)$ & $\!0.016492(10)$   & $\!-0.1237(30)$\\
$256$&   Th  & $\ \ 0.188235\dots$ & $\ 0.0116877\dots$ & $-0.08487\dots$ \\
\ \  &   N   & $0.18823(1)$ & $\!\!\!0.011690(7)$   & $\!-0.0833(29)$ \\
\end{tabular}
\end{ruledtabular}
\end{table}   

We then made long runs with $N_{\rm sam}=10^7$ for $L=16$ for both $\rho=1/2$ and $1/4$,
with the results displayed in 
Table~\ref{t4}. In this case the numerical results clearly confirm
the theoretical prediction for the skewness: very accurately for $\rho=1/4$,
and within reasonable error bars for $\rho=1/2$.

\begin{table}
\caption{\label{t4}
For systems with PBC, $L=16$ sites, and $\rho$ as specified, $\langle a_L\rangle \equiv 
\langle A \rangle/L$ is average current activity 
[$\,$normalized as in
 Eq.~(\protect{\ref{eq:a-jp}})$\,$]; $\Delta \equiv [\langle a_L^2\rangle -
\langle a_L\rangle^2\,]^{1/2}$; 'Th' stands for results from theory 
[$\,$Eqs.~(\protect{\ref{eq:momn1p}})--(\protect{\ref{eq:momn3p}})$\,$], and 'N' for
results of numerical simulations with $N_{\rm sam}=10^7$, $N_{\rm set}=10$ (see text).
}
\vskip 0.2cm
\begin{ruledtabular}
\begin{tabular}{@{}cccc}
 Type  & $\langle a_L \rangle$ & $\Delta$ & Skew  \\
\hline\noalign{\smallskip}
\multicolumn{4}{c}{$\rho=1/2$}\\
\hline\noalign{\smallskip}
Th  &   $\ \ 0.266667\dots$  & $\ \ 0.0623610\dots$ & $\,0.005140\dots$ \\
N   & $0.26668(1)$ & $0.062350(8)$   & $0.00515(72)$\\
\hline\noalign{\smallskip}
\multicolumn{4}{c}{$\rho=1/4$}\\
\hline\noalign{\smallskip}
Th  & $1/5$ & $\ \ 0.0443203\dots$ & $\,-0.455600\dots$  \\
N   & $0.20002(1)$   & $0.044313(4)$     & $-0.45578(30)$\\
\end{tabular}
\end{ruledtabular}
\end{table}   

As a consequence of Eq.~(\ref{eq:spbc1}), for $\rho \neq 1/2$
the skewness of the PDFs for PBC tends to grow, in absolute value, as $|\rho-1/2|$ 
increases. For system sizes $L$ large enough, so that the set of (discrete) 
allowed activity values can be reasonably well represented as a continuum,
one can approximate the respective PDF by a Gaussian with a skew-inducing
perturbation:
\begin{equation}
P(x)=\left[1+a\left(\frac{x^\prime}{\sigma}\right)^3\right]\,G(x^\prime,\sigma)\ ,
\label{eq:modgaus}
\end{equation}
where $x^\prime \equiv x-x_0$, $ G(x^\prime,\sigma)$ is a Gaussian curve
centered at $x=x_0$, with width $\sigma$; $a$, $x_0$, and $\sigma$
are adjustable parameters, with the proviso $|a| \ll 1$.      

\begin{figure}
{\centering \resizebox*{3.3in}{!}{\includegraphics*{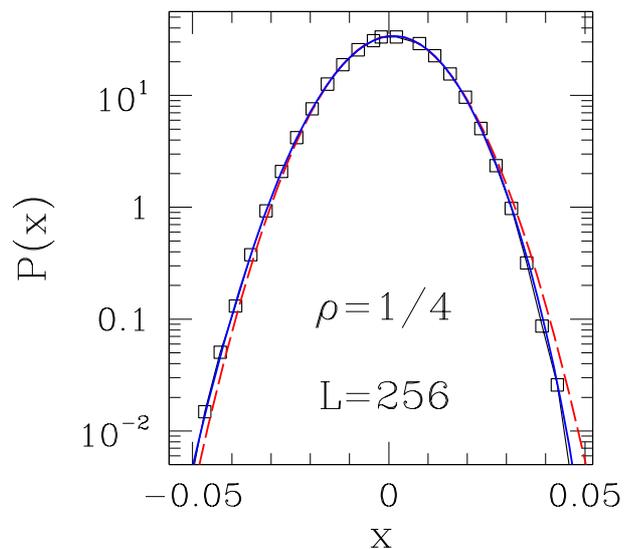}}}
\caption{(Color online) 
PDF for the reduced variable $x \equiv (A-\langle A \rangle)/L$
for system with PBC, $\rho=1/4$, $L=256$ [$\,$see
Eq.~(\protect{\ref{eq:a-jp}})$\,$]. 
The points are from numerical simulations, $N_{\rm sam}=10^6$ samples. 
The full (blue) line is the perturbed Gaussian distribution given by
Eq.~(\protect{\ref{eq:modgaus}}), adjusted to the numerical data. 
The dashed (red) line is a pure
Gaussian fit to the same data, and is shown for reference. 
} 
\label{fig:rho025}
\end{figure}

In Fig.~\ref{fig:rho025} we show the results of fitting data for $L=256$,
$\rho=1/4$ to Eq.~(\ref{eq:modgaus}). The adjusted estimate of $a=-0.0120(25)$
corresponds to $S=-0.072(15)$, which compares favorably
with the values given in Table~\ref{t3}, albeit with a
somewhat large uncertainty. The adjusted width $\sigma=0.01171(3)$
is in very good agreement with the $\Delta$ values (both from theory and
from numerics) given in the same Table.

\begin{figure}
{\centering \resizebox*{3.3in}{!}{\includegraphics*{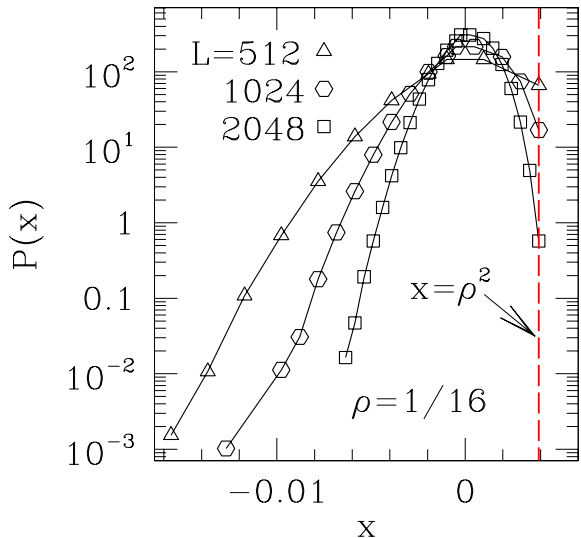}}}
\caption{(Color online) 
PDFs for the reduced variable $x \equiv (A-\langle A \rangle)/L$
for systems with PBC, $\rho=1/16$, and sizes $L$ as specified [$\,$see
Eq.~(\protect{\ref{eq:a-jp}})$\,$]. 
The points are from numerical simulations, $N_{\rm sam}=10^6$ samples. 
The dashed (red) line corresponds to the maximum allowed activity (apart from
small corrections of order $L^{-1}$, see text). 
} 
\label{fig:rho116vl}
\end{figure}

For a graphic illustration of large skewness with PBC (similar to the
effect shown in Fig.~\ref{fig:a05vl} for open BC with
$\alpha \ll 1$, $\beta \approx 1$), one needs $|\rho -1/2|$ large, 
as well as 
large $L$. 
In the low-density regime, the maximum allowed activity is $A_{\rm max}=M$,  
$M$ being the number of particles in the system.
One has, from Eq.~(\ref{eq:momn1p}), $\langle A \rangle =M(1-\rho)\left(1+
{\cal O}(L^{-1})\right)$, thus the maximum allowed value of $x=(A-\langle A \rangle)/L$ 
is $\rho^2\left(1+{\cal O}(L^{-1})\right)$. This is illustrated in Fig.~\ref{fig:rho116vl},
for $\rho=1/16$ and various values of (large) $L$, where the sharp cutoff in the
forward end of the PDF is evident. For the curves shown, the calculated values of
the skewness are $S=-0.58963\dots$, $-0.41258\dots$, and $-0.29023\dots$,
respectively for $L=512$, $1024$, and $2048$ 
[$\,$see Eqs.~(\ref{eq:momn1p})--(\ref{eq:momn3p})$\,$].
The low-density regime with PBC is thus the counterpart of the small $\alpha$, large $\beta$
example for open BC of section~\ref{sec:nr-obc}. Note
 that the skewness here has the opposite sign to that case.

\begin{figure}
{\centering \resizebox*{3.3in}{!}{\includegraphics*{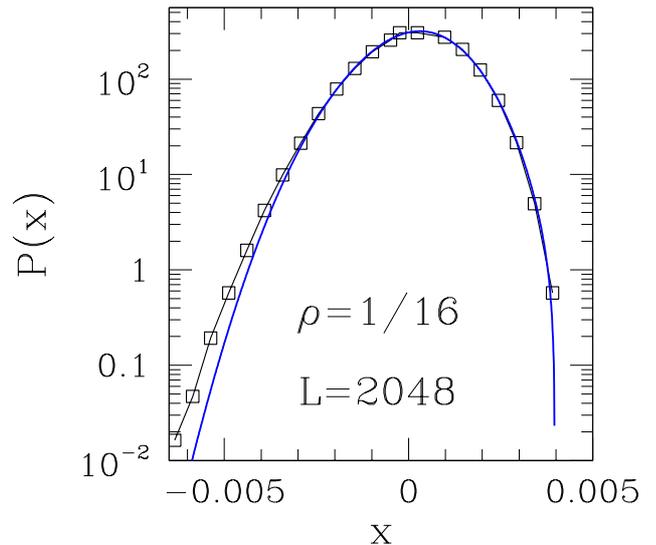}}}
\caption{(Color online) 
PDF for the reduced variable $x \equiv (A-\langle A \rangle)/L$
for system with PBC, $\rho=1/16$, $L=2048$ [$\,$see
Eq.~(\protect{\ref{eq:a-jp}})$\,$]. 
The points are from numerical simulations, $N_{\rm sam}=10^6$ samples. 
The full (blue) line is the perturbed Gaussian distribution given by
Eq.~(\protect{\ref{eq:modgaus}}), adjusted to the numerical data. 
} 
\label{fig:rho116}
\end{figure}

We considered the system with $\rho=1/16$, 
$L=2048$ in further detail.
Fig.~\ref{fig:rho116} shows the results of
numerical simulations, as well as their best fit to Eq.~(\ref{eq:modgaus}).
The adjusted value of $a=-0.043(9)$ corresponds to $S=-0.27(6)$, in broad
agreement with the theoretical prediction. However, it can be
seen that the perturbative scheme of  Eq.~(\ref{eq:modgaus}) cannot
properly account for the long tail at negative $x$. Although we have
checked that the tail still varies as $\sim \exp[-b(x-x_0)^2]$,
it is not possible to find a single set of parameters in Eq.~(\ref{eq:modgaus})
which optimizes the fit to the latter region, without seriously compromising
the description of  data away from the tail.

In Ref.~\onlinecite{szl03}, an approach similar to the one described
above was used to calculate current 
activity PDFs for one-dimensional flow of extended objects with exclusion, on a
ring. Those were compared to results of numerical simulations of the 
same quantity, with very good agreement (see their Figure 1).
Note that their system of
extended objects corresponds to an effective lattice of $L=35$ sites with
$M=15$ point particles. For this case Eqs.~(\ref{eq:momn1p})--(\ref{eq:momn3p})
give, e.g., $S=-0.015\dots$. So this is a regime where deviations from
Gaussianity are rather small for activity distributions, which is visually 
confirmed by the aspect of their figure referred to above.

\section{Discussion and Conclusions} 
\label{sec:conc}
The investigation of current fluctuations is usually carried out
by examining the total charge crossing a given bond, during a long time interval
in the steady state regime~\cite{dem95,dl98,bd06,kvo10,gv11,lm11,ess11}. 
Conversely, the snapshot nature of current activity 
means that samples are collected at a fixed time; in this case, the non-trivial
features arise when one considers the corresponding global quantity, i.e.,
the sum of contributions from all bonds in the system, at a given instant. 
When focusing on fluctuations of either quantity, the task for currents is made  
more involved by the need to subtract a background term which grows linearly
in time, and (depending on the specifics of the case), additional sublinear
terms as well. Current activity fluctuations, on the other hand,
can be adequately sampled by resorting to large enough systems, so that 
discrete-lattice effects are minimized. While, as emphasized earlier, these two
quantities are distinct in character, both can yield physical insights into the 
properties of flow with exclusion. 

The results of the numerical simulations, reported in Section~\ref{sec:nr},
provide unequivocal support to the predictions summarized in 
Eqs.~(\ref{eq:momn1})--(\ref{eq:momn3}) [$\,$for the cases where 
Eqs.~(\ref{eq:znsimpa}) and~(\ref{eq:znsimpb}) apply$\,$], 
and in Eqs.~(\ref{eq:momn1p})--(\ref{eq:momn3p}).
We thus collected further information, regarding points of the $\alpha-\beta$ phase 
diagram for open
BC not investigated in Section~\ref{sec:nr}, by 
applying  the general expression for the quantities $Z_L$, Eq.~(\ref{eq:za-bgen}),
directly to Eqs.~(\ref{eq:momn1})--(\ref{eq:momn3}). The results are as follows.

We focused on the large--$L$  dependence of the cumulants $C_n$, $n=1-3$, as reflected 
in the associated quantities $\Delta=C_2^{1/2}/(L+1)$ and $S=C_3/C_2^{3/2}$,
as well as on the sign of the latter. 

\begin{figure}
{\centering \resizebox*{3.3in}{!}{\includegraphics*{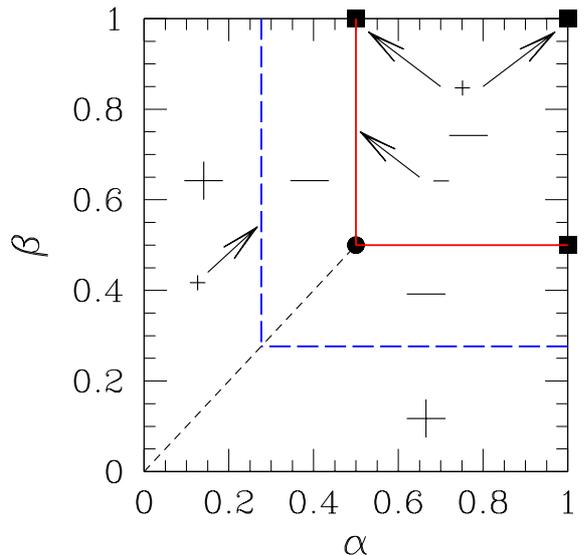}}}
\caption{(Color online) 
The sign of skewness $S$ in the various regions of the $\alpha-\beta$ phase diagram is shown.
The large--$L$ dependence, $|S| \propto L^{-x}$, is:
in the low-current phases [$\alpha <1/2$, $\beta$ and $\beta< 1/2$, $\alpha\,$]
$x=1/2$, except for the long-dashed (blue) lines, including their
extremes at $\beta=1$ and $\alpha=1$  $[\,x=3/2\,]$ (see text for
explanation of these lines).
In the high-current phase $\alpha$, $\beta>1/2$, $x=3/2$.
Full (red) lines separating high- and low-current phases: $x=3/2$.
Full squares: $x=5/2$. 
On the $\alpha=1$ and $\beta=1$ lines, $S$
has the same sign and $L$-dependence as in the respective adjacent regions,
except for the points marked by full squares.  
The circle marks $(\alpha,\beta)=(1/2,1/2)$
where $S \equiv 0$
[$\,$see Eq.~(\protect{\ref{eq:gfab05}})$\,$]. 
The short-dashed line is the coexistence line between high-and low-density phases.
Because of particle-hole duality, $S$ is the same for pairs of points
symmetric with respect to  the $\alpha=\beta$ line.} 
\label{fig:pd}
\end{figure}

From numerical evaluations of Eqs.~(\ref{eq:momn1})--(\ref{eq:momn3})
using Eq.~(\ref{eq:za-bgen}) we found $\Delta \propto L^{-1/2}$
everywhere, and strong evidence for
\begin{equation}
S \propto L^{-k_m-3/2}\ 
\label{eq:sa-bgen}
\end{equation}
in the limit of large $L$, where $k_m=-1$ for some extended regions of the
$\alpha-\beta$ phase diagram, $k_m=0$ for other regions, as well as special lines,
and $k_m=1$ for some special points (see Figure~\ref{fig:pd}). 
Eq.~(\ref{eq:sa-bgen}) corresponds to $C_3 \propto L^{-k_m}$ at large $L$. 
All this is consistent with results obtained  by using in
Eqs.~(\ref{eq:momn1})--(\ref{eq:momn3}), instead of Eq.~(\ref{eq:za-bgen})
for $Z_L$,
the asymptotic approximations to it provided by Refs.~\onlinecite{derr93}, 
\onlinecite{ds04,ds05} (see, e.g., Eqs.~(8) and ~(9) of Ref.~\onlinecite{ds05}).
This gives, for the low-current phases [$\,\alpha < 1/2$, $\beta$ (low density)
or $\beta < 1/2$, $\alpha$ (high density)$\,$] and the coexistence line
($\alpha=\beta < 1/2$) the results
\begin{eqnarray}
C_2 = L \mu (1-3\mu)+{\cal O}(1)\ ;\nonumber \\
\quad C_3 =L \mu (1-4\mu)(1-5\mu)+{\cal O}(1)\ ,
\label{eq:c2c3lc}
\end{eqnarray}
where $\mu= {\rm min}\,(\alpha,\beta)\,[1-{\rm min}\,(\alpha,\beta)]$.
Thus, in these cases, away from the dual lines $\mu=1/5$ the large--$L$
skewness is $S \propto L^{-1/2}$ and positive (negative) for $\mu$ less
(greater) than $1/5$. At $\mu=1/5$, corresponding to
${\rm min}\,(\alpha,\beta)=\frac{1}{2}-\frac{\sqrt{5}}{10}=0.27639\dots$,
the vanishing of the ${\cal O}(L)$ term in $C_3$ makes 
$|S| \leq {\cal O}(L^{-3/2})$. These agree with the signs and $L$--dependences
of $S$ from the numerical calculations using Eq.~(\ref{eq:za-bgen}),
gathered in Figure~\ref{fig:pd}. There the long-dashed lines are the
dual lines where $\mu=1/5$. They cross the $\alpha+\beta=1$ line at
$\alpha\beta=1/5$. At this point, the term in $C_3$ contributing to
$S \propto L^{-1/2}$ indeed vanishes, as anticipated in Eq.~(\ref{eq:c3}).
In fact, 
Eqs.~(\ref{eq:c2c3lc}) reproduce Eqs.~(\ref{eq:c2}) and~(\ref{eq:c3}), 
everywhere along the $\alpha+\beta=1$ line, including $\alpha=\beta=1/2$
where, from Section~\ref{sec:a+b1}, we know that $S$ vanishes.

For $\alpha$, $\beta$ inside the high current phase boundaries, 
using the asymptotic $Z_L$ (Eq.~(8) of
Ref.~\onlinecite{ds05}) in Eqs.~(\ref{eq:momn1})--(\ref{eq:momn3})
gives $S \propto L^{-3/2}$, in agreement with the results obtained
using the full form, Eq.~(\ref{eq:za-bgen}). Furthermore, the full form shows 
that near $(\alpha,\beta)=(\frac{1}{2},
\frac{1}{2})$, $S >0$ for small $L$, crossing over to negative values as
$L$ increases. For example, at $\alpha=\beta=0.55$, $S$ goes through zero 
for $L \approx 300$. The $L^{-3/2}$ dependence involves an exception to the
normal subdominance of injection and ejection contributions:  
both the approximate and exact
procedures show that without the injection and ejection contribution
[$\,$i.e., considering the $\langle {A^\prime}^n\rangle$ of 
Eqs.~(\ref{eq:momn1})--(\ref{eq:momn3})$\,$], $S$ would have been
${\cal O}(L^{-5/2})$; an associated relationship to cumulants of $A^\prime$
helps to explain the result $S \propto L^{-5/2}$ at the special boundary
point $\alpha=\beta=1$ [$\,$see Eq.~(\ref{eq:skab1})$\,$].

It is remarkable that the asymptotic approximations for $Z_L$ are sufficiently
accurate, when inserted in Eqs.~(\ref{eq:momn1})--(\ref{eq:momn3}),
to give the correct asymptotics for the cumulants and skewness,
given the high orders of cancellations in $1/L$ expansions often occurring.
Direct calculation of activity cumulants using asymptotic PDFs given at the
end of Section~\ref{sec:actdist2} are less accurate: 
for example,  the Laplace approximation mentioned there
gives, inside the low current phases,
$S \propto L^{-3/2}$, but does not give the crossover near the long-dashed
(blue) lines of Fig.~\ref{fig:pd}.

Correspondence with select results for standard current studies is as follows.
In Ref.~\onlinecite{kvo10}, current statistics were considered for open BC.
At large $L$, for the PDF of total charge fluctuations, ${\widetilde Q}=
Q(T)-\langle Q(T)\rangle$, accumulated up to time $T$ they found 
critical scaling at $\alpha=\beta=1/2$ with $T^{1/2z}\,P_T({\widetilde Q})=
f({\widetilde Q}/T^{1/2z})$, $z=3/2$; elsewhere on the 
$\alpha+\beta=1$ line, scaling of the PDF was found to be Gaussian.
Although their results are strictly not comparable to ours,
their Figure 4 suggests that, e. g., for $L=256$ critical scaling
would only persist up to times  $T \lesssim 3,000$. Whether or not some such signature
would show up in the (fixed-time) statistics of current activity, for times other
than those used in the present work, is an open question.

Secondly, in Ref.~\onlinecite{lm11} it was found that, at the special point
$\alpha=\beta=1$, $C_3$ approaches a small, finite (negative) value, as $L \to \infty$.  
On the other hand, their $C_2 \propto L^{-1/2}$. Thus, the standard skewness~\cite{numrec}
would become very large for large $L$. Here, in contrast, the skewness of
the current activity approaches zero as $L \to \infty$ everywhere. Furthermore,  it is
positive at the above-mentioned special point. 
This implies that any relationship of 
current fluctuations to those of the activity would involve subtle features.

\begin{acknowledgments}
The authors thank Fabian Essler for helpful discussions.
S.L.A.d.Q. thanks the Rudolf Peierls Centre for Theoretical Physics,
Oxford, where most of this work was carried out, for the hospitality,
and CAPES for funding his visit. The research of S.L.A.d.Q. is financed 
by the Brazilian agencies CAPES (Grant No. 0940-10-0),  
CNPq  (Grant No. 302924/2009-4), and FAPERJ (Grant No. E-26/101.572/2010).
\end{acknowledgments}
 
\appendix

\section{Transfer matrix results for $\alpha+\beta=1$}
\label{sec:app1} 

The evaluation of Eq.~(\ref{eq:genfunc2}) using the representation in which $T$
is diagonal yields
\begin{equation}
\langle e^{\lambda\,A}\rangle = a_+\,\mu_+^{L-1}+a_-\,\mu_-^{L-1}\ ,
\label{eq:ev12b}
\end{equation}
where
\begin{equation}
\mu_\pm=\mu_\pm(\lambda) =\frac{1}{2}\left[1 \pm 
\sqrt{1+4\alpha\beta\gamma}\right]\quad (\gamma \equiv e^\lambda-1) 
\label{eq:mupmb}
\end{equation}
and
\begin{equation}
a_\pm(\lambda)=[(\alpha\,e^{\lambda\beta}+\beta)\mu_\pm+\alpha\beta\gamma]\,
\frac{[\mu_\pm+\beta(e^{\lambda\alpha}-1)]}{[(2\mu_\pm-1)\mu_\pm]}\ .
\label{eq:apm}
\end{equation}
The first three derivatives of $\ln \langle e^{\lambda\,A}\rangle$ at $\lambda=0$
then provide the following leading cumulants (exploiting $\mu_+(0)=1$,
$\mu_-(0)=0$):
\begin{eqnarray}
C_1=(L+1)\,\alpha\beta\hskip5.3truecm \nonumber\\
C_2=L\,\alpha\beta-(3L+1)\,(\alpha\beta)^2\hskip3.5truecm \nonumber\\
C_3=L\,\alpha\beta-(9L-1)\,(\alpha\beta)^2+4(5L-1)\,(\alpha\beta)^3\ .\quad\
\label{eq:excum}
\end{eqnarray}
These results generalize Eqs.~(\ref{eq:c1})--(\ref{eq:c3}) and show that
Eqs.~(\ref{eq:c1}), (\ref{eq:c2}), and~(\ref{eq:c3}) apply for $L \gg 1$,
$\alpha\beta/(1-3\alpha\beta)$, $\alpha\beta/(1-5\alpha\beta)$ respectively. 

The exact cumulants given above can alternatively be obtained from the forms
provided for general $\alpha$, $\beta$ in Section~\ref{sec:gen_mom},
taking Eq.~(\ref{eq:znsimpa}) into account.

\section{$\langle {A^\prime}^3\rangle$, $\langle A^3\rangle$ for general $\alpha$, 
$\beta$}
\label{sec:app2} 

We first consider the product of bond terms
\begin{eqnarray}
{A^\prime}^3= \sum_{\ell=1}^{L-1} (1-\zeta_\ell)\,\zeta_{\ell+1}\,\sum_{m=1}^{L-1} 
(1-\zeta_m)\,\zeta_{m+1}\,\sum_{n=1}^{L-1} (1-\zeta_n)\,\zeta_{n+1}\nonumber\\ 
\equiv \sum_{\ell m n} T_{\ell m n}\ .\hskip4truecm
\label{eq:a3def}
\end{eqnarray}
$(\ell,\,m,\,n)$ identifies a particular term in the triple sum, and corresponds
to a point in an $(L-1)\times (L-1)\times (L-1)$ cube in $3$-dimensional
$\,\ell\,m\,n\,$ space. 

We need to distinguish and enumerate the following cases ($a\,$)--($f\,$), of which
($a\,$)--($e\,$) involve repeated site labels:
\smallskip\par\noindent
\ ($a\,$)\ $(\ell\,m\,n)=(\ell,\ell,\ell)$, $1\leq \ell \leq L-1$: $N_a=L-1$ points
(on the diagonal of the cube). 
$T_{\ell\,\ell\,\ell}=(1-\zeta_\ell)^3\,\zeta_{\ell+1}^3=
(1-\zeta_\ell)\,\zeta_{\ell+1}$.
\smallskip\par\noindent
\ ($b\,$)\ $(\ell\,m\,n)=(\ell,\ell,\ell+1)$, $1\leq \ell \leq L-2$: $L-2$ points
(on subdiagonal line). $T_{\ell\,\ell\,\ell+1}=(1-\zeta_\ell)^2\,\zeta_{\ell+1}^2\,
(1-\zeta_{\ell+1})\,\zeta_{\ell+2} =0$~.
\par\noindent This is one example out of $6$ lines of the type
$(\ell\,m\,n)=(\ell,\ell,\ell\pm 1)$ and cyclic variations, containing altogether
$N_b=6(N-2)$ points each giving zero contribution.
\smallskip\par\noindent
\ ($c\,$)\ $(\ell\,m\,n)=(\ell,\ell \pm 1,\ell \mp1)$, ($\ell$ varying;
$\ell$, $m$, $n$ all different): with cyclic variations,
there are $6$ lines containing points of this type, altogether making $N_c=6(L-3)$
points, all with $T_{\ell\,m\,n}=0$.
\smallskip\par\noindent
\ ($d\,$)\ $(\ell\,m\,m)$: as $\ell$, $m$ vary, this corresponds to a plane slicing 
the cube. It contains $(L-1)^2$ points of which $L-1$ are on the line
$(\ell\,m\,n)=(\ell,\ell,\ell)$ specified in ($a\,$), and $2(L-2)$ are on lines
$(\ell,\ell\pm 1,\ell\pm 1)$ of the type specified in ($b\,$). So there are
$(L-1)^2-(L-1+2(L-2))$ "new" points associated with the plane. Allowing for cyclic
variations there are three such planes, containing altogether
$N_d=3(L-2)(L-3)$ "new" points, each with 
$T_{\ell\,m\,m}=(1-\zeta_\ell)\,\zeta_{\ell+1}\,
(1-\zeta_m)\,\zeta_{m+1}$ $(m \neq \ell$, $\ell\pm 1)$.
\smallskip\par\noindent
\ ($e\,$)\ $(\ell,\,m,\,m \pm 1)$  with $\ell$, $m$ varying (and cyclic
variations). There are six planes of this type. 
Discounting points already accounted for, the
new planes contain a total of $N_e=6(L-3)(L-4)$ "new" points. at each of
which $T_{\ell\,m\,n}$ vanishes.
\smallskip\par\noindent
\ ($f\,$)\ The remaining points $(\ell\,m\,n)$ have no pair of coordinates
equal or differing by $\pm 1$, so the associated $T_{\ell\,m\,n}$ has no shared
site labels. The total number of such points is $N_f=(L-1)^3-(N_a+N_b+N_c+
N_d+N_e)=(L-3)(L-4)(L-5)$.
\smallskip\par
The average of the addition of the non-zero contributions to ${A^\prime}^3$
from ($a\,$), ($d\,$), ($f\,$) gives
\begin{eqnarray}
\langle {A^\prime}^3 \rangle=N_a\,\langle (1-\zeta_\ell)\,\zeta_{\ell+1}\rangle+
\hskip3truecm\nonumber \\
+N_d\,\langle (1-\zeta_\ell)\,\zeta_{\ell+1}(1-\zeta_m)\,\zeta_{m+1}\rangle+ 
\hskip2truecm\nonumber \\
+N_f\,\langle 
(1-\zeta_\ell)\,\zeta_{\ell+1}\,(1-\zeta_m)\,\zeta_{m+1}\,(1-\zeta_n)\,
\zeta_{n+1}\rangle\ .
\hskip1truecm
\label{eq:a3pf}
\end{eqnarray}
Here no pair from $\ell\,m\,n$ are equal or differ by $\pm 1$,
so we can reduce Eq.~(\ref{eq:a3pf}) 
using Eqs.~(\ref{eq:exn1}),~(\ref{eq:exn2}), 
and their generalization for this case
\begin{equation}
\langle 
(1-\zeta_\ell)\,\zeta_{\ell+1}\,(1-\zeta_m)\,\zeta_{m+1}\,
(1-\zeta_n)\,\zeta_{n+1}\rangle 
=\frac{Z_{L-3}}{Z_L}\ .
\label{eq:exn3}
\end{equation}
We find
\begin{eqnarray}
\langle {A^\prime}^3 \rangle =(L-3)(L-4)(L-5)\,\frac{Z_{L-3}}{Z_L}+\hskip2truecm
\nonumber\\ +3(L-2)(L-3)\,\frac{Z_{L-2}}{Z_L}+(L-1)\,\frac{Z_{L-1}}{Z_L}\ .
\hskip.5truecm
\label{eq:momn3b}
\end{eqnarray} 
The generalization for inclusion of injection and ejection contributions
in the third moment of the activity involves similar
steps to those for the second moment, given in Eq.~(\ref{eq:a2av}). We first write
\begin{equation}
A^3=T_0+3T_1+3T_2+T_3\ ,
\label{eq:a3vsap3}
\end{equation}
where
\begin{equation}
 T_n=(A^\prime)^n\,\left(\alpha\,\zeta_1+\beta\,(1-\zeta_L)\right)^{3-n}\ .
\label{eq:a3vsap3b}
\end{equation}
Using $(\alpha\,\zeta_1)^m=\alpha^m\,\zeta_1$,  
$(\alpha\,\zeta_1)^m\,(1-\zeta_1)\,\zeta_2=0$ ($m\geq 1$) and similarly for
$\beta\,(1-\zeta_L)$, one finds
\begin{eqnarray}
\langle T_0 \rangle =(\alpha^2+\beta^2)\,\frac{Z_{L-1}}{Z_L}+
3(\alpha+\beta)\,\frac{Z_{L-2}}{Z_L}\ ,\hskip1truecm\nonumber\\
\langle T_1 \rangle =(L-2)(\alpha+\beta)\,\frac{Z_{L-2}}{Z_L}+
2(L-3)\,\frac{Z_{L-3}}{Z_L}\ ,\hskip1truecm\nonumber\\
\langle T_2 \rangle =2(L-2)\,\frac{Z_{L-2}}{Z_L}+
2(L-3)(L-4)\,\frac{Z_{L-3}}{Z_L}\ .\hskip1truecm
\label{eq:t012}
\end{eqnarray}
\smallskip\par
With $\langle T_3\rangle=\langle {A^\prime}^3\rangle$ from Eq.~(\ref{eq:momn3b}),
\begin{eqnarray}
\langle A^3 \rangle =(L-1)(L-2)(L-3)\,\frac{Z_{L-3}}{Z_L}+\hskip2truecm
\nonumber\\ +3(L-1)(L-2+\alpha+\beta)
\,\frac{Z_{L-2}}{Z_L}+\hskip1.5truecm \nonumber\\ 
+(L-1+\alpha^2+\beta^2)\,\frac{Z_{L-1}}{Z_L}\ .\hskip2.78truecm
\label{eq:momn3c}
\end{eqnarray}

\section{$\langle A^3 \rangle$ for PBC}
\label{sec:app3}

Along similar lines to those followed in Appendix~\ref{sec:app2},
we consider the product:
\begin{equation}
A^3= \sum_{\ell=1}^L (1-\zeta_\ell)\,\zeta_{\ell+1}\,\sum_{m=1}^L 
(1-\zeta_m)\,\zeta_{m+1}\,\sum_{n=1}^L (1-\zeta_n)\,\zeta_{n+1}\ , 
\label{eq:a3pdef}
\end{equation}
with $\zeta_{L+1} \equiv \zeta_1$. 

The cases involving site labels are the same set ($a$)-($f$) described in
Appendix~\ref{sec:app2}; only the enumerations vary. Recalling Eq.~(\ref{eq:a3pf}), 
one has now, adding up the nonzero contributions:
\begin{eqnarray}
\langle A^3 \rangle=N_a^p\,\langle (1-\zeta_\ell)\,\zeta_{\ell+1}\rangle+
\hskip3truecm\nonumber \\
+N_d^p\,\langle (1-\zeta_\ell)\,\zeta_{\ell+1}(1-\zeta_m)\,\zeta_{m+1}\rangle+ 
\hskip2truecm\nonumber \\
+N_f^p\,\langle 
(1-\zeta_\ell)\,\zeta_{\ell+1}\,(1-\zeta_m)\,\zeta_{m+1}\,
(1-\zeta_n)\,\zeta_{n+1}\rangle\ .
\hskip1truecm
\label{eq:a3fp}
\end{eqnarray}

The numerical coefficients are as follows: 
\smallskip\par\noindent
\ ($a\,$)\ $(\ell\,m\,n)=(\ell,\ell,\ell)$, $1\leq \ell \leq L$: $N_a^p=L$ points.
\par\noindent
\ ($d\,$)\ $(\ell\,m\,m)$: $N_d^p=3L(L-3)$ points.
\par\noindent
\ ($f\,$)\ The points $(\ell\,m\,n)$ which have no pair of coordinates
equal or differing by $\pm 1$: $N_f^p=L(L-4)(L-5)$ points.
\smallskip\par
One has:
\begin{equation}
\langle (1-\zeta_\ell)\,\zeta_{\ell+1}\rangle= \frac{_{L-2}C_{M-1}}{_LC_M}\ ;
\label{eq:nap}
\end{equation}

\begin{equation}
\langle (1-\zeta_\ell)\,\zeta_{\ell+1}(1-\zeta_m)\,\zeta_{m+1}\rangle
= \frac{_{L-4}C_{M-2}}{_LC_M}\ ;
\label{eq:ndp}
\end{equation}

\begin{equation}
\langle 
(1-\zeta_\ell)\,\zeta_{\ell+1}\,(1-\zeta_m)\,\zeta_{m+1}\,
(1-\zeta_n)\,\zeta_{n+1}\rangle\
= \frac{_{L-6}C_{M-3}}{_LC_M}\ .
\label{eq:nfp}
\end{equation}
So, recalling Eq.~(\ref{eq:andef}),
\begin{equation}
\langle A^3 \rangle=a_0\,(1+3 a_1+a_1\,a_2)\ .
\label{eq:momn3pap}
\end{equation}

\end{document}